\documentstyle[12pt,epsfig,axodraw]{article}
\newcommand {\ignore}[1]{}
\def\smallfrac#1#2{{\textstyle{#1 \over #2}}}
\def\lsim{\:\raisebox{-0.5ex}{$\stackrel{\textstyle<}{\sim}$}\:} 
\def\gsim{\:\raisebox{-0.5ex}{$\stackrel{\textstyle>}{\sim}$}\:} 
\def\fig#1{{Fig. (\ref{#1})}}
\def\smallfrac#1#2{{\textstyle{#1 \over #2}}}
\def\N{$\cal N$ }
\def\slash#1{#1\!\!\! /}

\def\rp{$R_p \hspace{-1em}/\;\:$ }
\def\Eq#1{{Eq. (\ref{#1})}}
\def\eq#1{{eq. (\ref{#1})}}

\def\be{\begin{equation}}
\def\ee{\end{equation}}
\def\bear{\be\begin{array}}
\def\eear{\end{array}\ee} 
\def\bea{\begin{eqnarray}} 
\def\eea{\end{eqnarray}}

\def\vb#1{\vbox to #1 pt{}}
\def\beqa{\begin{eqnarray}}
\def\eeqa{\end{eqnarray}}
\def\ni{\noindent}
\def\beq{\begin{equation}}
\def\eeq{\end{equation}}
\def\ba{\begin{array}}
\def\ea{\end{array}}
\def\ovl{\overline}
\def\ds{\displaystyle}

\def\npb#1#2#3{{\it Nucl.\ Phys.\ }{\bf B #1} (#2) #3}
\def\plb#1#2#3{{\it Phys.\ Lett.\ }{\bf B #1} (#2) #3}
\def\prd#1#2#3{{\it Phys.\ Rev.\ }{\bf D #1} (#2) #3}

\def\hepph#1{{\tt hep-ph/#1}}
\def\nm{\hbox{$\nu_\mu$ }}
\def\nt{\hbox{$\nu_\tau$ }}
\def\21{$SU(2) \otimes U(1)$}

\def\half{{\textstyle{1 \over 2}}} 
 
\def\quarter{{\textstyle{1 \over 4}}} 
 
\def\eighth{{\textstyle{1 \over 8}}} 
\def\sqrthalf{{\textstyle{1 \over \sqrt{2}}}} 
\def\bold#1{\setbox0=\hbox{$#1$} 
     \kern-.025em\copy0\kern-\wd0 
     \kern.05em\copy0\kern-\wd0 
     \kern-.025em\raise.0433em\box0 } 
\begin{document} 
\begin{titlepage} 
\begin{flushright}
hep-ph/0004115 \\ 
FTUV-00-0412\\ 
IFIC/00-24\\ 
%%\today
\end{flushright} 
\vspace*{3mm} 
\begin{center}  
  \textbf{\Large Neutrino Masses and Mixings from Supersymmetry
    with Bilinear R--Parity Violation:}
 \textbf{\Large A Theory for Solar and Atmospheric Neutrino Oscillations}\\[10mm]

{M. Hirsch${}^1$,  M. A. D\'{\i}az${}^2$,
W. Porod${}^1$, J. C. Rom\~ao${}^3$ and 
J. W. F. Valle${}^1$ } 
\vspace{0.3cm}\\ 

{\it $^1$ Departament de F\'\i sica Te\`orica, IFIC--CSIC,
          Universitat de Val\`encia\\
          46100 Burjassot, Val\`encia, Spain \\ }
{\it $^2$ Facultad de F\'\i sica, Universidad Cat\'olica de Chile\\ 
          Av. Vicu\~na Mackenna 4860, Santiago, Chile\\}
{\it $^3$ Departamento de F\'\i sica, Instituto Superior T\'ecnico\\
          Av. Rovisco Pais 1, $\:\:$ 1049-001 Lisboa, Portugal \\}

\end{center}

%%\vspace{5mm}

\begin{abstract} 
  \small The simplest unified extension of the Minimal Supersymmetric
  Standard Model with bi-linear R--Parity violation naturally predicts
  a hierarchical neutrino mass spectrum, in which one neutrino
  acquires mass by mixing with neutralinos, while the other two get
  mass radiatively.  We have performed a full one-loop calculation of
  the neutralino-neutrino mass matrix in the bi-linear \rp MSSM,
  taking special care to achieve a manifestly gauge invariant
  calculation.  Moreover we have performed the renormalization of the
  heaviest neutrino, needed in order to get meaningful results.  The
  atmospheric mass scale and maximal mixing angle arise from
  tree-level physics, while solar neutrino scale and oscillations
  follow from calculable one-loop corrections. If universal
  supergravity assumptions are made on the soft-supersymmetry breaking
  terms then the atmospheric scale is calculable as a function of a
  single \rp violating parameter by the renormalization group
  evolution due to the non-zero bottom quark Yukawa coupling. The
  solar neutrino problem must be accounted for by the small mixing
  angle (SMA) MSW solution.  If these assumptions are relaxed then one
  can implement large mixing angle solutions, either MSW or just-so.
  The theory predicts the lightest supersymmetic particle (LSP) decay
  to be observable at high-energy colliders, despite the smallness of
  neutrino masses indicated by experiment.  This provides an
  independent way to test this solution of the atmospheric and solar
  neutrino anomalies.

\end{abstract} 
 
\end{titlepage}

%\tableofcontents

\newpage

\setcounter{page}{1} 

\section{Introduction}

The high statistics data by the SuperKamiokande collaboration
\cite{Fukuda:1998mi} has confirmed the deficit of atmospheric muon
neutrinos, especially at small zenith angles, opening a new era in
neutrino physics.  On the other hand the persistent disagreement
between solar neutrino data and theoretical expectations has been a
long-standing problem in physics~\cite{Smy:1999tt}.  Altogether these
constitute the only solid evidence we now have in favour of physics
beyond the present standard model, providing a strong hint for
neutrino conversion.  Although massless neutrino
conversions~\cite{Valle:1987gv} can be sizeable in matter, and may
even provide alternative solutions of the neutrino anomalies
~\cite{Fornengo:1999zp}, it is fair to say that the simplest
interpretation of the present data is in terms of massive neutrino
oscillations.
Taking for granted such an interpretation, the present data do
provide an important clue on the pattern of neutrino masses and
mixing.
The atmospheric data indicate \nm to \nt flavour oscillations with
maximal mixing ~\cite{atm99}, while the solar data can be accounted
for in terms of either small (SMA) and large (LMA) mixing MSW
solutions~\cite{MSW99}, as well as through vacuum or \emph{ just-so}
solutions \cite{Barger:1999sm}. A large mixing among $\nu_{\tau}$ and
$\nu_e$ is excluded both by the atmospheric data and by reactor 
data on neutrino oscillations ~\cite{chooz}.
There has indeed been an avalanche \cite{avalanche} of papers trying
to address this issue in the framework of unified models adopting
\emph{ad hoc} texture structures for the Yukawa couplings.

Here we propose an alternative approach to describe the structure of
lepton mixing which accounts for the atmospheric and solar neutrino
anomalies~\cite{epsrad} based on the simplest extension of minimal
supergravity with bi-linear R--Parity violation \cite{e3others}. The
particles underlying the mechanism of neutrino mass generation are the
neutral supersymmetric partners of the Standard Model gauge and Higgs
bosons which have mass at the weak-scale and are thus accessible to
accelerators.

Our model breaks lepton number and therefore necessarily generates
non-zero Majorana neutrino masses~\cite{Schechter:1980gr}. At
tree-level only one of the neutrinos picks up a mass by mixing with
neutralinos~\cite{Ross:1985yg}, leaving the other two neutrinos
massless ~\cite{ProjectiveMassMatrix}. While this can explain the
atmospheric neutrino problem, to reconcile it with the solar neutrino
data requires going beyond the tree-level approximation. This is the
purpose of the present paper. Here we improve the work of ref.
\cite{Hempfling} by performing a full one-loop calculation of the
neutrino mass matrix and also update the discussion in the light of
the recent global fits of solar and atmospheric neutrino data. This
can also be used to improve the discussion given in
\cite{Romao:1992ex} where  the tree approximation was assumed.  For
simplified analyses including only the atmospheric neutrino problem in
the tree-level approximation see ref.~\cite{Bednyakov:1998cx} and a
number of papers in ref.~\cite{epsnuothers}.

We have performed a full one-loop calculation of the
neutralino-neutrino mass matrix in the bi-linear \rp MSSM, showing
that, in order to explain the solar and atmospheric neutrino data, it
is necessary and sufficient to work at the one-loop level, provided
one performs the renormalization of the heaviest neutrino. In contrast
to all existing papers \cite{Hempfling,epsnuothers}, we have taken
special care to verify the gauge invariance of the calculation, thus
refining the approximate approaches so far used in the literature. We
find that if the soft-supersymmetry breaking terms are universal at
the unification scale then only the small mixing angle (SMA) MSW
solution to the solar neutrino problem exists.  On the other hand if
these assumptions are relaxed then one can implement large mixing
angle solutions, either MSW or just-so.

Bilinear R-parity breaking supersymmetry has been extensively
discussed in the literature~\cite{epsrad}. It is motivated on the one
hand by the fact that it provides an effective truncation of models
where R--parity breaks \emph{spontaneously} by singlet sneutrino vevs
around the weak scale~\cite{SBRpV}. Moreover, they allow for the
radiative breaking of R-parity, opening also new ways to unify Gauge
and Yukawa couplings~\cite{Diaz:1999wz} and with a potentially
slightly lower prediction for $\alpha_s$~\cite{Diaz:1999is}. For
recent papers on phenomenological implications of these models see
ref.~\cite{rphen00,chargedhiggs,Hirsch:1999kc}. If present at the
fundamental level tri--linear breaking of R--parity will always imply
bi-linear breaking at some level, as a result of the renormalization
group evolution. In contrast, bi-linear breaking may exist in the
absence of tri--linear, as would be the case if it arises
spontaneously.

This paper is organized as follows. In sections \ref{TheModel},
\ref{ThePot} and \ref{ewbreaking} we describe the model, the
minimization of the scalar potential and the radiative breaking of the
electroweak symmetry. In section \ref{TreeLevelMasses} the tree level
masses and mixings are described, while the contributions to the one
loop mass matrix and the gauge invariance issue are studied in section
\ref{OneLoopMassMatrix}.  Finally the neutrino masses and mixings are
discussed in section \ref{OneLoopMasses} where we show our results for
solar and atmospheric oscillation parameters. The more technical
questions regarding the mass matrices, couplings and one loop results
as well as further details of gauge invariance are given in the
appendices. We also briefly discuss how, despite the smallness of
neutrino masses indicated by experiment, the theory can lead to
observable \rp phenomena at high-energy accelerators.
 
\section{The Superpotential and the Soft Breaking Terms}
\label{TheModel}

Using the conventions of refs.~\cite{chargedhiggs,HabKaneGun} we
introduce the model by specifying the superpotential, which includes
BRpV \cite{epsrad} in three generations. It is given by
\begin{equation}  
W=\varepsilon_{ab}\left[ 
 h_U^{ij}\widehat Q_i^a\widehat U_j\widehat H_u^b 
+h_D^{ij}\widehat Q_i^b\widehat D_j\widehat H_d^a 
+h_E^{ij}\widehat L_i^b\widehat R_j\widehat H_d^a 
-\mu\widehat H_d^a\widehat H_u^b 
+\epsilon_i\widehat L_i^a\widehat H_u^b\right] 
\label{eq:Wsuppot} 
\end{equation} 
where the couplings $h_U$, $h_D$ and $h_E$ are $3\times 3$ Yukawa
matrices and $\mu$ and $\epsilon_i$ are parameters with units of mass.
The bilinear term in eq.~(\ref{eq:Wsuppot}) violates lepton number in
addition to R--Parity.

Supersymmetry breaking is parameterized with a set of soft supersymmetry 
breaking terms. In the MSSM these are given by
\begin{eqnarray} 
{\cal L}_{soft}^{MSSM}&=& 
M_Q^{ij2}\widetilde Q^{a*}_i\widetilde Q^a_j+M_U^{ij2} 
\widetilde U_i\widetilde U^*_j+M_D^{ij2}\widetilde D_i 
\widetilde D^*_j+M_L^{ij2}\widetilde L^{a*}_i\widetilde L^a_j+ 
M_R^{ij2}\widetilde R_i\widetilde R^*_j \nonumber\\ 
&&\!\!\!\!+m_{H_d}^2 H^{a*}_d H^a_d+m_{H_u}^2 H^{a*}_u H^a_u- 
\left[\half M_s\lambda_s\lambda_s+\half M\lambda\lambda 
+\half M'\lambda'\lambda'+h.c.\right]\label{eq:Vsoft} \\ 
&&\!\!\!\!+\varepsilon_{ab}\left[ 
A_U^{ij}\widetilde Q_i^a\widetilde U_j H_u^b 
+A_D^{ij}\widetilde Q_i^b\widetilde D_j H_d^a 
+A_E^{ij}\widetilde L_i^b\widetilde R_j H_d^a 
-B\mu H_d^a H_u^b\right] 
\,.\nonumber 
\end{eqnarray} 
In addition to the MSSM soft SUSY breaking terms in ${\cal
  L}_{soft}^{MSSM}$ the BRpV model contains the following extra term
\begin{equation}
V_{soft}^{BRpV} = - B_i\epsilon_i\varepsilon_{ab}\widetilde 
L_i^a H_u^b\,,
\label{softBRpV}
\end{equation}
where the $B_i$ have units of mass. In what follows, we neglect
intergenerational mixing in the soft terms in eq.~(\ref{eq:Vsoft}).

The electroweak symmetry is broken when the two Higgs doublets $H_d$ and 
$H_u$, and the neutral component of the slepton doublets $\widetilde L^1_i$ 
acquire vacuum expectation values. We introduce the notation: 
\begin{equation} 
H_d={{H^0_d}\choose{H^-_d}}\,,\qquad 
H_u={{H^+_u}\choose{H^0_u}}\,,\qquad
\widetilde L_i={{\tilde L^0_i}\choose{\tilde\ell^-_i}}\,,
\label{eq:shiftdoub} 
\end{equation} 
where we shift the neutral fields with non--zero vevs as
\begin{equation}
H_d^0\equiv{1\over{\sqrt{2}}}[\sigma^0_d+v_d+i\varphi^0_d]\,,\quad
H_u^0\equiv{1\over{\sqrt{2}}}[\sigma^0_u+v_u+i\varphi^0_u]\,,\quad
\tilde{L}_i^0\equiv{1\over{\sqrt{2}}}[\tilde\nu^R_i+v_i+i\tilde\nu^I_i]\,.
\label{shiftedfields}
\end{equation}
Note that the $W$ boson acquires a mass $m_W^2=\quarter g^2v^2$, where
$v^2\equiv v_d^2 + v_u^2 + v_1^2+ v_2^2+ v_3^2 \simeq (246 \;
\rm{GeV})^2$.  We introduce the following notation in spherical
coordinates for the vacuum expectation values:
\begin{eqnarray} 
v_d&=&v\sin\theta_1\sin\theta_2\sin\theta_3\cos\beta\nonumber\\
v_u&=&v\sin\theta_1\sin\theta_2\sin\theta_3\sin\beta\nonumber\\ 
v_3&=&v\sin\theta_1\sin\theta_2\cos\theta_3\label{eq:vevs}\\
v_2&=&v\sin\theta_1\cos\theta_2\nonumber\\
v_1&=&v\cos\theta_1\nonumber
\end{eqnarray} 
which preserves the MSSM definition $\tan\beta=v_u/v_d$. In the MSSM
limit, where $\epsilon_i=v_i=0$, the angles $\theta_i$ are equal to
$\pi/2$.  In addition to the above MSSM parameters, our model contains
nine new parameters, $\epsilon_i$, $v_i$ and $B_i$. The three vevs are
determined by the one--loop tadpole equations, and we will assume
universality of the $B$--terms, $B=B_i$ at the unification scale.
Therefore, the only new and free parameters can be chosen as the
$\epsilon_i$.

\section{The Scalar Potential}
\label{ThePot}

The electroweak symmetry is broken when the Higgs and lepton fields
acquire non--zero vevs. These are calculated via the minimization of
the effective potential or, in the diagramatic method, via the tadpole
equations.  The full scalar potential at tree level is
\begin{equation} 
V_{total}^0  = \sum_i \left| { \partial W \over \partial z_i} \right|^2 
        + V_D + V_{soft}^{MSSM} + V_{soft}^{BRpV}
\label{V}
\end{equation} 
where $z_i$ is any one of the scalar fields in the superpotential in
eq.~(\ref{eq:Wsuppot}), $V_D$ are the $D$-terms, and $V_{soft}^{BRpV}$ is
given in eq.~(\ref{softBRpV}). 

The tree level scalar potential contains the following linear terms 
\begin{equation}
V_{linear}^0=t_d^0\sigma^0_d+t_u^0\sigma^0_u+t_1^0\tilde\nu^R_1
+t_2^0\tilde\nu^R_2+t_3^0\tilde\nu^R_3\,,
\label{eq:Vlinear}
\end{equation}
where the different $t^0$ are the tadpoles at tree level. They are given by
\begin{eqnarray}
t_d^0&=&\Big(m_{H_d}^2+\mu^2\Big)v_d+v_dD-\mu\Big(Bv_u+v_i\epsilon_i\Big)
\nonumber\\
t_u^0&=&-B\mu v_d+\Big(m_{H_u}^2+\mu^2\Big)v_u-v_uD+v_iB_i\epsilon_i
+v_u\epsilon^2
\nonumber\\
t_1^0&=&v_1D+\epsilon_1\Big(-\mu v_d+v_uB_1+v_i\epsilon_i\Big)+\half\Big(
v_iM^2_{Li1}+M^2_{L 1i}v_i\Big)
\label{eq:tadpoles}\\
t_2^0&=&v_2D+\epsilon_2\Big(-\mu v_d+v_uB_2+v_i\epsilon_i\Big)+\half\Big(
v_iM^2_{Li2}+M^2_{L2i}v_i\Big)
\nonumber\\
t_3^0&=&v_3D+\epsilon_3\Big(-\mu v_d+v_uB_3+v_i\epsilon_i\Big)
+\half\Big(v_iM^2_{Li3}+M^2_{L3i}v_i\Big)
\nonumber
\end{eqnarray}
where we have defined
$D=\eighth(g^2+g'^2)(v_1^2+v_2^2+v_3^2+v_d^2-v_u^2)$ and
$\epsilon^2=\epsilon_1^2+\epsilon_2^2+\epsilon_3^2$. A repeated index
$i$ in eq.~(\ref{eq:tadpoles}) implies summation over $i=1,2,3$. The
five tree level tadpoles $t_{\alpha}^0$ are equal to zero at the
minimum of the tree level potential, and from there one can determine
the five tree level vacuum expectation values.

It is well known that in order to find reliable results for the electroweak
symmetry breaking it is necessary to include the one--loop radiative 
corrections. The full scalar potential at one loop level, called effective 
potential, is
\beq
V_{total}  = V_{total}^0 + V_{RC}
\eeq
where $V_{RC}$ include the quantum corrections. In this paper we use the 
diagramatic method, which incorporates the radiative corrections through 
the one--loop corrected tadpole equations. The one loop tadpoles are
\begin{equation}
t_{\alpha}=t^0_{\alpha} -\delta t^{\overline{DR}}_{\alpha}
+T_{\alpha}(Q)=t^0_{\alpha} +\widetilde T^{\overline{DR}} _{\alpha}(Q)
\label{tadpoles}
\end{equation}
where $\alpha=d,u,1,2,3$ and $\widetilde T^{\overline{DR}} _{\alpha}(Q)
\equiv-\delta t^{\overline{MS}}_{\alpha}+T_{\alpha}(Q)$ are the finite one 
loop tadpoles. At the minimum of the potential we have $t_{\alpha}=0$, and 
the vevs calculated from these equations are the renormalized vevs.

Neglecting intergenerational mixing in the soft masses, the five
tadpole equations can be conveniently written in matrix form as
\begin{equation}
\left[t_u^0,t_d^0,t_1^0,t_2^0,t_3^0 \right]^T=
{\bf{M}^2_{tad}}\left[v_u,v_d,v_1,v_2,v_3\right]^T
\label{tadpoleMat}
\end{equation}
where the matrix ${\bf{M}^2_{tad}}$ is given by 
\begin{equation}
{\bf{M}^2_{tad}}=\!
\left[\matrix{
m_{H_d}^2\!+\!\mu^2\!+\!D & -B\mu & -\mu\epsilon_1 & -\mu\epsilon_2 & 
-\mu\epsilon_3 \cr
-B\mu & \!\!\!\!m_{H_u}^2\!+\!\mu^2\!+\epsilon^2\!-\!D & B_1\epsilon_1 & 
B_2\epsilon_2 & B_3\epsilon_3 \cr
-\mu\epsilon_1 & B_1\epsilon_1 & \!\!\!\!M_{L_1}^2\!+\epsilon_1^2\!+\!D & 
\epsilon_1\epsilon_2 & \epsilon_1\epsilon_3 \cr
-\mu\epsilon_2 & B_2\epsilon_2 & \epsilon_1\epsilon_2 & 
\!\!\!\!M_{L_2}^2\!+\epsilon_2^2\!+\!D & \epsilon_2\epsilon_3 \cr
-\mu\epsilon_3 & B_3\epsilon_3 & \epsilon_1\epsilon_3 & 
\epsilon_2\epsilon_3 & \!\!\!\!M_{L_3}^2\!+\epsilon_3^2\!+\!D
}\right]
\label{Mtad}
\end{equation}
and depends on the vevs only through the $D$ term defined above. 

In order to have approximate solutions for the tree level vevs,
consider the following rotation among the $H_d$ and lepton
superfields:
\begin{equation}
{\bf{M'}^2_{tad}}={\bf R}{\bf{M}^2_{tad}}{\bf R^{-1}}
\label{rotMtad}
\end{equation}
where the rotation ${\bf R}$ can be split as
\begin{equation}
{\bf R}=\left[\matrix{c_3 & 0 &   0  &   0  & -s_3 \cr
                       0  & 1 &   0  &   0  &   0  \cr
                       0  & 0 &   1  &   0  &   0  \cr
                       0  & 0 &   0  &   1  &   0  \cr
                      s_3 & 0 &   0  &   0  &  c_3 }\right]
  \times\left[\matrix{c_2 & 0 &   0  & -s_2 &   0  \cr
                       0  & 1 &   0  &   0  &   0  \cr
                       0  & 0 &   1  &   0  &   0  \cr
                      s_2 & 0 &   0  &  c_2 &   0  \cr
                       0  & 0 &   0  &   0  &   1  }\right]
  \times\left[\matrix{c_1 & 0 & -s_1 &   0  &   0  \cr
                       0  & 1 &   0  &   0  &   0  \cr
                      s_1 & 0 &  c_1 &   0  &   0  \cr
                       0  & 0 &   0  &   1  &   0  \cr
                       0  & 0 &   0  &   0  &   1  }\right]\,.
\label{rotationR}
\end{equation}
where the three angles are defined as
\begin{eqnarray}
&&c_1={{\mu}\over{\mu'}}\,,\qquad\,\, s_1={{\epsilon_1}\over{\mu'}}\,,
\qquad\,\,\mu'=\sqrt{\mu^2+\epsilon_1^2}\,,\nonumber\\
&&c_2={{\mu'}\over{\mu''}}\,,\qquad\, s_2={{\epsilon_2}\over{\mu''}}\,,
\qquad\,\mu''=\sqrt{\mu'^2+\epsilon_2^2}\,,\label{angles}\\
&&c_3={{\mu''}\over{\mu'''}}\,,\qquad s_3={{\epsilon_3}\over{\mu'''}}\,,
\qquad\mu'''=\sqrt{\mu''^2+\epsilon_3^2}\,.\nonumber
\end{eqnarray}
It is clear that this rotation ${\bf R}$ leaves the $D$ term
invariant.  The rotated vevs are given by
\begin{equation}
\left[v'_u,v'_d,v'_1,v'_2,v'_3\right]^T=
{\bf{R}}\left[v_u,v_d,v_1,v_2,v_3\right]^T\,,
\label{rotatedvevs}
\end{equation}
and under the assumption that $v'_1,v'_2,v'_3\ll v$, these three small
vevs have the approximate solution
\begin{eqnarray}
&&v'_1\approx-{{\mu\epsilon_1}\over{M'^2_{L_1}+D}}\left[
{{m_{H_d}^2-M_{L_1}^2}\over{\mu'\mu'''}}v'_d+{{B_1-B}\over{\mu'}}v'_u
\right]\,,\nonumber\\\nonumber\\
&&v'_2\approx-{{\mu'\epsilon_2}\over{M'^2_{L_2}+D}}\left[
{{m'^2_{H_d}-M_{L_2}^2}\over{\mu''\mu'''}}v'_d+{{B_2-B'}\over{\mu''}}v'_u
\right]\,,\label{vevsPapprox}\\\nonumber\\
&&v'_3\approx-{{\mu''\epsilon_3}\over{M'^2_{L_3}+D}}\left[
{{m''^2_{H_d}-M_{L_3}^2}\over{\mu'''^2}}v'_d+{{B_3-B''}\over{\mu'''}}v'_u
\right]\,,\nonumber
\end{eqnarray}
where we have defined the following rotated soft terms:
\begin{eqnarray}
&&\!\!\!\!\!\!\!\!\!\!\!\!\!\!\!\!\!m'^2_{H_d}=
{{m_{H_d}^2\mu^2+M_{L_1}^2\epsilon_1^2}\over{\mu'^2}}\,,\quad
m''^2_{H_d}={{m'^2_{H_d}\mu'^2+M_{L_2}^2\epsilon_2^2}\over{\mu''^2}}
\,,\quad
m'''^2_{H_d}={{m''^2_{H_d}\mu''^2+M_{L_3}^2\epsilon_3^2}\over{\mu'''^2}}
\,,\nonumber\\
&&\!\!\!\!\!\!\!\!\!\!\!\!\!\!\!\!\!B'=
{{B\mu^2+B_1\epsilon_1^2}\over{\mu'^2}}\,,\qquad\quad\,\,\,
B''={{B'\mu'^2+B_2\epsilon_2^2}\over{\mu''^2}}\,,\qquad\quad
B'''={{B''\mu''^2+B_3\epsilon_3^2}\over{\mu'''^2}}\,,
\label{newsoft}\\
&&\!\!\!\!\!\!\!\!\!\!\!\!\!\!\!\!\!M'^2_{L_1}=
{{m_{H_d}^2\epsilon_1^2+M_{L_1}^2\mu^2}\over{\mu'^2}}\,,\quad
M'^2_{L_2}={{m'^2_{H_d}\epsilon_2^2+M_{L_2}^2\mu'^2}\over{\mu''^2}}\,,
\quad
M'^2_{L_3}={{m''^2_{H_d}\epsilon_3^2+M_{L_3}^2\mu''^2}\over{\mu'''^2}}\,.
\nonumber
\end{eqnarray}
The approximation $v'_1,v'_2,v'_3\ll v$ is justified in SUGRA models with
universality of soft masses at the weak scale, as shown in the next section.

\section{Radiative Breaking of the Electroweak Symmetry}
\label{ewbreaking}

It was demonstrated in ref.~\cite{epsrad} that BRpV can be succesfully
embedded into SUGRA with universal boundary conditions at the
unification scale, and with a radiatively broken electroweak symmetry.
At $Q = M_{U}$ we assume the standard minimal supergravity unification
assumptions,
\begin{eqnarray}
&&A_t=A_b=A_{\tau}\equiv A\:,\nonumber\\
&&B=B_i=A-1 \:,\nonumber\\
%%&&m_{H_d}^2=m_{H_u}^2=M_{L_i}^2=M_{R_i}^2=m_0^2 \:,\label{Ucondit}\\
%%&&M_{Q_i}^2=M_{U_i}^2=M_{D_i}^2=m_0^2\:,\nonumber\\
&&m_{H_d}^2=m_{H_u}^2=M_{L_i}^2=M_{R_i}^2 = M_{Q_i}^2=M_{U_i}^2=M_{D_i}^2
=m_0^2\:,\label{Ucondit}\\
&&M_3=M_2=M_1=M_{1/2}\,.\nonumber
\end{eqnarray}
We run the RGE's from the unification scale $M_{U}\sim2\times10^{16}$ GeV 
down to the weak scale, giving random values to the fundamental parameters 
at the unification scale:
\beq
\begin{array}{ccccc}
10^{-2} & \leq &{h^2_t}_{U} / 4\pi & \leq&1 \cr
10^{-5} & \leq &{h^2_b}_{U} / 4\pi & \leq&1 \cr
-3&\leq& a_0 \equiv A/m_0&\leq&3 \cr
0&\leq&\mu^2_{U}/m_0^2&\leq&10 \cr
0&\leq&M_{1/2}/m_0&\leq&5 \cr
\end{array}
\eeq
The Yukawa couplings are determined by requiring that three
eigenvalues of the chargino/charged-lepton mass matrix corrrespond to
the experimentally measured tau, muon, and electron masses
\footnote{For the case of large tree-level neutrino mass one must note
  that the lepton Yukawa couplings are no longer related to the lepton
  masses via the simple relations valid in the Standard Model. Since
  charginos mix with charged leptons, the Yukawa couplings depend also
  on the parameters of the chargino sector. For the case of interest
  here (light \nt mass fixed by the atmospheric scale) this correction
  is less important.}

As in the MSSM, the electroweak symmetry is broken because the large
value of the top quark mass drives the Higgs mass parameter
$m_{H_U}^2$ to negative values at the weak scale via its RGE
\cite{Ibanez:1982fr}. In the rotated basis, the parameter $\mu'''^2$ is
determined at one loop by
\begin{equation}
\mu'''^2=-{1\over 2}\left[m_Z^2-\widetilde A_{ZZ}(m_Z^2)\right]+
{{\left(m'''^2_{H_d}+\widetilde T^{\overline{DR}}_{v'_d}\,\right)-
\left(m_{H_u}^2+\widetilde T^{\overline{DR}}_{v'_u}\,\right)t'^2_{\beta}
}\over{t'^2_{\beta}-1}}
\label{muppp2}
\end{equation}
where $t'_{\beta}=v'_u/v'_d$ is defined in the rotated basis and is
analogous to $\tan\beta$ in eq.~(\ref{eq:vevs}) defined in the
original basis. The finite $\overline{DR}$ Z-boson self energy is
$\widetilde A_{ZZ}(m_Z^2)$, and the one--loop tadpoles
$T^{\overline{DR}}_{v'_d}$ and $T^{\overline{DR}}_{v'_u}$ are obtained
by applying to the original tadpoles in eq.~(\ref{tadpoles}) the
rotation $\bf R$ defined in eq.~(\ref{rotationR}).  The radiative
breaking of the electroweak symmetry is valid in the BRpV model in the
usual way: the large value of the top quark Yukawa coupling drives the
parameter $m_{H_U}^2$ to negative values, breaking the symmetry of the
scalar potential.

As we will see a radiative mechanism is also responsible for the
smallness of the neutrino masses in models with universality of soft
mass parameters at the unification scale. The relevant parameters are
the bilinear mass parameters $B$ and $B_i$, the Higgs mass parameter
$m_{H_d}^2$, and the slepton mass parameters $M_{L_i}^2$.

The RGE's for the $B$ parameters are
\begin{eqnarray}
&&{{dB}\over{dt}}={1\over{8\pi^2}}\left(
3h_t^2A_t+3h_b^2A_b+h_{\tau}^2A_{\tau}+3g_2^2M_2+\textstyle{3\over 5}
g_1^2M_1\right)\nonumber\\
&&{{dB_3}\over{dt}}={1\over{8\pi^2}}\left(
3h_t^2A_t+h_{\tau}^2A_{\tau}+3g_2^2M_2+\textstyle{3\over 5}
g_1^2M_1\right)\label{RGEBs}\\
&&{{dB_2}\over{dt}}={{dB_1}\over{dt}}={1\over{8\pi^2}}\left(
3h_t^2A_t+3g_2^2M_2+\textstyle{3\over 5}
g_1^2M_1\right)\,,\nonumber
\end{eqnarray}
where we do not write the effect of Yukawa couplings of the first two 
generations. Similarly, the RGE for the down-type Higgs mass is
\begin{equation}
{{dm_{H_d}^2}\over{dt}}={1\over{8\pi^2}}\Big(3h_b^2X_b+
h_{\tau}^2X_{\tau}-3g_2^2M_2^2-\textstyle{3\over 5}g_1^2M_1^2\Big)\,,
\label{RGEmHd}
\end{equation}
and the RGE's for the slepton mass parameters are
\begin{eqnarray}
&&{{dM_{L_3}^2}\over{dt}}={1\over{8\pi^2}}\Big(h_{\tau}^2X_{\tau}
-3g_2^2M_2^2-\textstyle{3\over 5}g_1^2M_1^2\Big)\nonumber\\
&&{{dM_{L_2}^2}\over{dt}}={{dM_{L_1}^2}\over{dt}}=-{1\over{8\pi^2}}\Big(
3g_2^2M_2^2+\textstyle{3\over 5}g_1^2M_1^2\Big)\,,
\label{RGEmsl}
\end{eqnarray}
where $X_b=m_{H_d}^2+M_{Q_3}^2+M_{D_3}^2+A_b^2$ and 
$X_{\tau}=m_{H_d}^2+M_{L_3}^2+M_{R_3}^2+A_{\tau}^2$.

With the aid of these RGE's we can find an approximate expression for
the slepton vev's in the rotated basis $v'_i$, given in
eq.~(\ref{vevsPapprox}).  The relevant soft term differences, defined
as $\Delta B_i\equiv B_i-B$ and $\Delta m_i^2\equiv
M_{L_i}^2-m_{H_d}^2$, are approximated by
\begin{eqnarray}
&&\Delta B_3={1\over{8\pi^2}}\Big(3h_b^2A_b\Big)
\ln{{M_{U}}\over{m_{weak}}}
\nonumber\\
&&\Delta B_2=\Delta B_1={1\over{8\pi^2}}\Big(3h_b^2A_b+
h_{\tau}^2A_{\tau}\Big)\ln{{M_{U}}\over{m_{weak}}}
\label{DBiapprox}
\end{eqnarray}
for the $B$ terms, and by 
\begin{eqnarray}
&&\Delta m_3^2={1\over{8\pi^2}}\Big(3h_b^2X_b\Big)
\ln{{M_{U}}\over{m_{weak}}}
\nonumber\\
&&\Delta m_2^2=\Delta m_1^2={1\over{8\pi^2}}
\Big(3h_b^2X_b+h_{\tau}^2X_{\tau}\Big)\ln{{M_{U}}\over{m_{weak}}}
\label{Dm2approx}
\end{eqnarray}
for the mass squared terms. This way, if we assume that
$\epsilon_i\ll\mu$ we can neglect the rotations in
eq.~(\ref{vevsPapprox}) and we find
\begin{equation}
v'_i\approx{{v_d\epsilon_i/\mu}\over{M^2_{L_i}+D}}
\Big(\Delta m^2_i-t_{\beta}\mu\Delta B_i\Big)
\label{vpiapprox}
\end{equation}
which give us an approximate expression for the sneutrino vev's $v'_i$
in the basis where the $\epsilon_i$ terms are absent from the
superpotential. In a model with unified universal boundary conditions
on the soft SUSY breaking terms (SUGRA case, for short) the $v'_i$ are
calculable in terms of the renormalization group evolution due to the
non-zero bottom quark Yukawa coupling.  We should stress here that for
our subsequent numerical calculation we solve the tadpole equations
\emph{exactly}.

The symmetry of the neutralino/neutrino mass matrix implies that only
one neutrino acquires a tree level mass, and the other two remain
massless~\cite{ProjectiveMassMatrix} (see next section). The massive
neutrino will have the largest component along $\tau$, $\mu$ or $e$ if
the largest vev is $v'_3$, $v'_2$, or $v'_1$ respectively. On the
other hand, the most obvious difference between the third generation
sneutrino vev and the first two generations is in the extra
contribution from $h_{\tau}$ to $\Delta B_i$ in eq.~(\ref{DBiapprox})
and to $\Delta m_i^2$ in eq.~(\ref{Dm2approx}) for the first two
generations. Due to the tau lepton contribution, $\Delta B_1$ and
$\Delta B_2$ are larger than $\Delta B_3$, and similarly for the
$\Delta m^2_i$, specially if $\tan\beta\gg 1$. However, we have
checked that it is possible without fine-tuning the parameters in an
unnatural way to arrange for the heaviest of the neutrinos to be an
equal mixture of \nm and $\nu_{\tau}$ as needed in order to obtain an
explanation of the atmospheric neutrino anomaly. That this is possible
can be understood by noticing that there can be a cancellation between
the $\Delta B$ and $\Delta m^2$ terms in eq.~(\ref{vpiapprox}) for
$v'_1$ and $v'_2$.  

\section{Tree Level Neutrino Masses and Mixings}
\label{TreeLevelMasses}

Here we discuss the tree level structure of neutrino masses and
mixings. For a complete discussion of the fermion mass matrices in
this model see Appendix~\ref{MassMatrices}. \footnote{In our notation
  the four component Majorana neutral fermions are obtained from the
  two component via the relation
$$
\chi_i^0=\left(\matrix{ F_i^0\cr \vb{18} \overline{F_i^0}\cr} \:.
\right) \nonumber
$$}
In the basis $\psi^{0T}=
(-i\lambda',-i\lambda^3,\widetilde{H}_d^1,\widetilde{H}_u^2, \nu_{e},
\nu_{\mu}, \nu_{\tau} )$ the neutral fermion mass matrix ${\bold M}_N$
is given by
\beq
{\bold M}_N=\left[  
\begin{array}{cc}  
{\cal M}_{\chi^0}& m^T \cr
\vb{20}
m & 0 \cr
\end{array}
\right]
\eeq
where
\beq
{\cal M}_{\chi^0}\hskip -2pt=\hskip -4pt \left[ \hskip -7pt 
\begin{array}{cccc}  
M_1 & 0 & -\frac 12g^{\prime }v_d & \frac 12g^{\prime }v_u \cr
\vb{12}   
0 & M_2 & \frac 12gv_d & -\frac 12gv_u \cr
\vb{12}   
-\frac 12g^{\prime }v_d & \frac 12gv_d & 0 & -\mu  \cr
\vb{12}
\frac 12g^{\prime }v_u & -\frac 12gv_u & -\mu & 0  \cr
\end{array}  
\hskip -6pt
\right] 
\eeq
is the standard MSSM neutralino mass matrix and
\beq
m=\left[  
\begin{array}{cccc}  
-\frac 12g^{\prime }v_1 & \frac 12gv_1 & 0 & \epsilon _1 \cr
\vb{18}
-\frac 12g^{\prime }v_2 & \frac 12gv_2 & 0 & \epsilon _2  \cr
\vb{18}
-\frac 12g^{\prime }v_3 & \frac 12gv_3 & 0 & \epsilon _3  \cr  
\end{array}  
\right] 
\eeq
characterizes the breaking of R-parity.  The mass matrix ${\bold M}_N$
is diagonalized by (see Appendix~\ref{MassMatrices})
\beq
{\cal  N}^*{\bold M}_N{\cal N}^{-1}={\rm diag}(m_{\chi^0_i},m_{\nu_j})
\label{chi0massmat}
\eeq
where $(i=1,\cdots,4)$ for the neutralinos, and $(j=1,\cdots,3)$ for
the neutrinos.  

We are interested in the case where the neutrino mass which is
determined at the tree level is small, since it will be determined in
order to account for the atmospheric neutrino anomaly.  The above form
for ${\bold M}_N$ is especially convenient in this case in order to
provide an approximate analytical discussion valid in the limit of
small \rp violation parameters. Indeed in this case we perform a
perturbative diagonalization of the neutral mass matrix, using the
method of \cite{Schechter:1982cv}, by defining~\cite{Hirsch:1999kc}
\beq
\xi = m \cdot {\cal M}_{\chi^0}^{-1}
\label{defxi}
\eeq
If the elements of this matrix satisfy
\beq
\forall \xi_{ij} \ll 1
\eeq
then one can use it as expansion parameter in order to find an
approximate solution for the mixing matrix ${\cal N}$.  Explicitly we
have
\begin{eqnarray}
\xi_{i1} &=& \frac{g' M_2 \mu}{2 det({\cal M}_{\chi^0})}\Lambda_i \cr
\vb{20}
\xi_{i2} &=& -\frac{g M_1 \mu}{2 det({\cal M}_{\chi^0})}\Lambda_i \cr
\vb{20}
\xi_{i3} &=& - \frac{\epsilon_i}{\mu} + 
          \frac{(g^2 M_1 + {g'}^2 M_2) v_u}
               {4 det({\cal M}_{\chi^0})}\Lambda_i \cr
\vb{20}
\xi_{i4} &=& - \frac{(g^2 M_1 + {g'}^2 M_2) v_d}
               {4 det({\cal M}_{\chi^0})}\Lambda_i
\label{xielementos}
\end{eqnarray}
where
\beq
\Lambda_i = \mu v_i + v_d \epsilon_i \propto  v'_i
\label{lambdai}
\eeq
are the alignment parameters. From \eq{xielementos} and \eq{lambdai}
one can see that $\xi=0$ in the MSSM limit where $\epsilon_i=0$,
$v_i=0$.  In leading order in $\xi$ the mixing matrix ${\cal N}$ is
given by,
\beq
{\cal N}^* \hskip -2pt =\hskip -2pt  \left(\hskip -1pt
\begin{array}{cc}
N^* & 0\\
0& V_\nu^T \end{array}
\hskip -1pt
\right)
\left(
\hskip -1pt
\begin{array}{cc}
1 -{1 \over 2} \xi^{\dagger} \xi& \xi^{\dagger} \\
-\xi &  1 -{1 \over 2} \xi \xi^\dagger
\end{array}
\hskip -1pt
\right) 
\eeq
The second matrix above block-diagonalizes the mass matrix ${\bold
  M}_N$ approximately to the form diag(${\cal M}_{\chi^0},m_{eff}$),
where
\beqa
m_{eff} &\hskip -3mm=\hskip -3mm& 
- m \cdot {\cal M}_{\chi^0}^{-1} m^T \cr
\vb{18}
&\hskip -3mm=\hskip -3mm& 
\frac{M_1 g^2 \!+\! M_2 {g'}^2}{4\, det({\cal M}_{\chi^0})} 
\left(\hskip -2mm \begin{array}{ccc}
\Lambda_e^2 
\hskip -1pt&\hskip -1pt
\Lambda_e \Lambda_\mu
\hskip -1pt&\hskip -1pt
\Lambda_e \Lambda_\tau \\
\Lambda_e \Lambda_\mu 
\hskip -1pt&\hskip -1pt
\Lambda_\mu^2
\hskip -1pt&\hskip -1pt
\Lambda_\mu \Lambda_\tau \\
\Lambda_e \Lambda_\tau 
\hskip -1pt&\hskip -1pt 
\Lambda_\mu \Lambda_\tau 
\hskip -1pt&\hskip -1pt
\Lambda_\tau^2
\end{array}\hskip -3mm \right)
\eeqa

\noindent
The sub-matrices $N$ and $V_{\nu}$ diagonalize ${\cal M}_{\chi^0}$ and
$m_{eff}$
\beq
N^{*}{\cal M}_{\chi^0} N^{\dagger} = {\rm diag}(m_{\chi^0_i}),
\eeq
\beq
V_{\nu}^T m_{eff} V_{\nu} = {\rm diag}(0,0,m_{\nu}),
\eeq
where 
\beq
\label{mnutree}
m_{\nu} = Tr(m_{eff}) = 
\frac{M_1 g^2 + M_2 {g'}^2}{4\, det({\cal M}_{\chi^0})} 
|{\vec \Lambda}|^2.
\eeq
Clearly, one neutrino acquires mass due to the projective nature of the
effective neutrino mass matrix $ m_{eff}$, a feature often encountered
in \rp models \cite{ProjectiveMassMatrix}. As a result one can rotate
away one of the three angles ~\cite{Schechter:1980gr} in the matrix
$V_{\nu}$, leading to~\cite{Schechter:1980bn}
\beq
V_{\nu}= 
\left(\begin{array}{ccc}
  1 &                0 &               0 \\
  0 &  \cos\theta_{23} & -\sin\theta_{23} \\
  0 &  \sin\theta_{23} & \cos\theta_{23} 
\end{array}\right) \times 
\left(\begin{array}{ccc}
  \cos\theta_{13} & 0 & -\sin\theta_{13} \\
                0 & 1 &               0 \\
  \sin\theta_{13} & 0 & \cos\theta_{13} 
\end{array}\right) ,
\eeq
where the mixing angles can be expressed in terms of the {\it
  alignment vector} ${\vec \Lambda}$ as follows:
\beq
\label{tetachooz}
\tan\theta_{13} = - \frac{\Lambda_e}
                   {(\Lambda_{\mu}^2+\Lambda_{\tau}^2)^{\frac{1}{2}}},
\eeq
\beq
\label{tetatm}
\tan\theta_{23} = \frac{\Lambda_{\mu}}{\Lambda_{\tau}}.
\eeq

\section{One Loop Neutrino Mass Matrix}
\label{OneLoopMassMatrix}

One--loop radiative corrections to the neutralino/neutrino mass matrix
in the BRpV model were calculated first in \cite{Hempfling}, working
in the t'Hooft--Feynman gauge ($\xi=1$). Our analysis improves the
previous work in that we check explicitly the gauge invariance using
the $R_{\xi}$ gauge. We use dimensional reduction to regularize the
divergences \cite{DR} and include all possible MSSM particles with
consistently determined mass spectra and couplings in the relevant
loops.

\subsection{Two--Point Function Renormalization}

We denote the sum of all one--loop graphs contributing to the 2-point
function as

\begin{center}
\begin{picture}(120,40)(0,22) % y_2 for equation position
\ArrowLine(5,25)(40,25)
\GCirc(60,25){20}{0.5}
\ArrowLine(80,25)(115,25)
\Text(25,34)[]{$F^0_i$}
\Text(100,34)[]{$F^0_j$}
\end{picture}
$
\equiv i\,{\bf\Sigma}_{FF}^{ij}(p)\,.
$
\end{center}
\vspace{5pt} \hfill \\
\noindent
The most general expression for the one--loop contribution to the 
unrenormalized neutralino/neutrino two--point function is
\begin{equation}
i\,{\bf \Sigma}_{FF}^{ij}(p)
\equiv i\left\{\vb{14}\slash{p}\left[\vb{12}P_L\Sigma^L_{ij}(p^2)+
P_R\Sigma^R_{ij}(p^2)\right]-\left[\vb{12}P_L\Pi^L_{ij}(p^2)+
P_R\Pi^R_{ij}(p^2)\right]\right\}
\end{equation}
where the indices $i$ and $j$ run from 1 to 7, $P_R=\half(1+\gamma_5)$
and $P_L=\half(1-\gamma_5)$ are the right and left projection
operators, and $p$ is the external four momenta. The functions
$\Sigma$ and $\Pi$ are unrenormalized self energies and depend on the
external momenta squared, $p^2$. The neutral fermions $F^0_i$ are a
mixture of weak eigenstate neutralinos and neutrinos and given by
\begin{equation}
F^0_i={\cal N}_{ij}\psi^0_j
\label{rotfields}
\end{equation}
where ${\cal N}$ is the $7\times7$ matrix that diagonalizes the
neutralino/neutrino mass matrix according to \eq{chi0massmat}.

The inverse propagator at one loop is obtained by adding to the tree
level propagator, this self energy previously renormalized with the
dimensional reduction $\overline{DR}$ scheme and denoted as
$\widetilde \Sigma$ and $\widetilde \Pi$. In the $\overline{DR}$
scheme, the counterterms cancel only the divergent pieces of the self
energies. In this way, they become finite and dependent on the
arbitrary scale $Q$. The tree level masses are promoted to running
masses in order to cancel the explicit scale dependence of the self
energies. Thus, the inverse propagator of the neutral fermion $F^0_i$
is
\begin{equation}
\Gamma^{(2)}_{FF}(p)=p_{\mu}\gamma^{\mu}-m_{F_i}(Q)+
\widetilde{\bf \Sigma}_{FF}^{ii}(p,Q)
\label{invprop}
\end{equation}
The physical pole mass is given by the zero of the inverse propagator, in
the limit where $p_{\mu}\gamma^{\mu}\rightarrow m_{F_i}$, and may be found 
using
\begin{equation}
\widetilde Z^{-1}_{F_i}\,\overline{u}(p)\Big[p_{\mu}\gamma^{\mu}-m_{F_i}
\Big]u(p)=\overline{u}(p)\Big[p_{\mu}\gamma^{\mu}-m_{F_i}(Q)+
\widetilde{\bf\Sigma}_{FF}^{ii}(p,Q)\Big]u(p)
\label{Zequation}
\end{equation}
where $u$ and $\overline{u}$ are two on-shell spinors, $m_{F_i}$ and
$m_{F_i}(Q)$ are the neutral fermion pole and running masses
respectively, and $\widetilde{\bf\Sigma}_{FF}^{ii}(p,Q)$ is the
renormalized two--point function in the $\overline{DR}$ scheme. The
quantity $\widetilde Z^{-1}_{F_i}$ corresponds to the finite ratio of
the infinite wave function renormalization constants in the
$\overline{DR}$ scheme and the on--shell scheme, and it accounts for
the fact that the residue of the $\overline{DR}$ propagator at the
pole is not one \cite{DKR}. The renormalized function
$\widetilde{\bf\Sigma}_{FF}^{ii}(p,Q)$ is calculated by subtracting
the pole terms proportional to the regulator of dimensional reduction
\beq
\ds \Delta=\frac{2}{4-d} -\gamma_E + \ln 4\pi
\eeq
where $\gamma_E$ is the Euler's constant and $d$ is the number of space--time
dimensions. In practice we have
\begin{equation}
\widetilde{\bf\Sigma}_{FF}^{ii}(p,Q)=
\left[{\bf\Sigma}_{FF}^{ii}(p)\right]_{\Delta=0}\,.
\label{delta0eq}
\end{equation}
Since $\overline u(p)\gamma_5u(p)=0$, the terms proportional to $\gamma_5$
in $\widetilde{\bf\Sigma}_{FF}^{ii}$ do not contribute. From 
eq.~(\ref{Zequation}) we find 
\begin{equation}
\Delta m_{F_i}\equiv m_{F_i}-m_{F_i}(Q)=\widetilde\Pi^V_{ii}(m_{F_i}^2)
-m_{F_i}\,\widetilde\Sigma^V_{ii}(m_{F_i}^2)\,,
\label{massdiff}
\end{equation}
where
\beq
\widetilde\Sigma^V=\half \left(\widetilde\Sigma^L+\widetilde\Sigma^R\right)
\,,\qquad
\widetilde\Pi^V=\half \left(\widetilde\Pi^L+\widetilde\Pi^R\right)\,,
\eeq
and the tilde implies renormalized self energies. A given set of input 
parameters in the neutralino/neutrino mass matrix defines the set of 
tree level running masses $m_{F_i}(Q)$, among them two massless and 
degenerate neutrinos. The one--loop renormalized masses $m_{F_i}$ are 
then found through eq.~(\ref{massdiff}), and the masslessness and 
degeneracy of the two lightest neutrinos is lifted.

The tree level masslessness of the lightest neutrinos implies an 
indetermination of the corresponding eigenvectors. In order to find the 
correct neutrino mixing angles we diagonalize the one--loop corrected 
neutralino/neutrino mass matrix. We define
\beq
M^{\rm pole}_{ij}= M^{\rm \overline{DR}}_{ij}(Q) + \Delta M_{ij}
\eeq
with
\beq
\Delta M_{ij}= \half 
\left[\widetilde\Pi^V_{ij}(m_i^2) + \widetilde\Pi^V_{ij}(m_j^2)\right] 
- \half 
\left[ m_{\chi^0_i} \widetilde\Sigma^V_{ij}(m_i^2)  +
m_{\chi^0_j} \widetilde\Sigma^V_{ij}(m_j^2) \right]\,,
\label{DeltaM}
\eeq
where the symmetrization is necessary to achieve gauge invariance. Of course,
the diagonal elements of $\Delta M_{ij}$ correspond to the difference 
between the pole and running masses defined in eq.~(\ref{massdiff}).

\subsection{Gauge Invariance}
\label{gauinv}

As explained in section 2.2, the one--loop corrected vacuum
expectation values are found by solving the one--loop corrected
tadpole equations in eq.~(\ref{tadpoles}). Of course, it is desirable
to work with gauge invariant vevs. In order to achieve the gauge
invariance of the $v_{\alpha}$'s, the one--loop tadpole $\widetilde
T^{\overline{DR}}_{\alpha}(Q)$ must be independent of the gauge
parameter $\xi$.  As it is shown in the appendix \ref{Ap:tadpoles}
the following set of tadpoles is gauge invariant:
\begin{center}
\begin{picture}(140,60)(0,26) % y_2 for equation position
\PhotonArc(20,50)(20,0,360){3}{21}
\DashLine(20,30)(20,0){4}
\Text(60,30)[]{$+$}
\Vertex(120,50){2}
  \Vertex(118.5,57.7){2}
 \Vertex(114.1,64.1){2}
  \Vertex(107.7,68.5){2}
\Vertex(100,70){2}
  \Vertex(92.3,68.5){2}
 \Vertex(85.9,64.1){2}
  \Vertex(81.5,57.7){2}
\Vertex(80,50){2}
  \Vertex(81.5,42.3){2}
 \Vertex(85.9,35.9){2}
  \Vertex(92.3,31.5){2}
\Vertex(100,30){2}
  \Vertex(107.7,31.5){2}
 \Vertex(114.1,35.9){2}
  \Vertex(118.5,42.3){2}
\DashLine(100,30)(100,0){4}
\Text(45,70)[]{$W$}
\Text(125,70)[]{$\eta^{\pm}$}
\Text(30,10)[]{$S'^0_{\alpha}$}
\Text(110,10)[]{$S'^0_{\alpha}$}
\end{picture}
$
=i\,[T_{\alpha}(Q)]^{W,\eta^{\pm}}
$
\end{center}
\vspace{5pt} \hfill \\
\noindent
where $S'^0_{\alpha}$ denote neutral scalar bosons in the weak basis
(see appendix A) $\eta$'s are the Fadeev-Popov ghosts.  A similar set
for the $Z$ gauge boson exists. Nevertheless, the tadpole with a
charged Goldstone boson in the loop introduces a gauge dependence that
cannot be canceled. For this reason, the Goldstone boson loops are
removed from the tadpoles $T_{\alpha}(Q)$ and introduced into the self
energies.  This in turn allows us to achieve the gauge invariance for
the two point functions, as well as for the vev's, as explained below.

Among the loops contributing to the self energies, consider for example 
the $W$--boson loop, which in the general $R_{\xi}$ gauge is

\begin{center}
\begin{picture}(120,40)(0,22) % y_2 for equation position
\ArrowLine(5,25)(40,25)
\ArrowArc(60,25)(20,180,0)
\PhotonArc(60,25)(20,0,180){3}{10.5}
\ArrowLine(80,25)(115,25)
\Text(25,34)[]{$F^0_i$}
\Text(100,34)[]{$F^0_j$}
\Text(75,50)[]{$W$}
\Text(75,0)[]{$F^+_k$}
\end{picture}
$
=i\,(\slash{p}\,{\bf\Sigma}_{ij}^V-\Pi^V_{ij})^W+...
$
\end{center}
\vspace{5pt} \hfill \\
\noindent
where the dots indicate terms proportional to $\gamma_5$ which are irrelevant
for us, $F^+_k$ are charged fermions resulting from mixing between charginos
and charged leptons, and 
\beqa
(\Sigma^V_{ij})^W
&\hskip -5pt=\hskip -5pt& 
-\frac{1}{16\pi^2}\, \sum_{k=1}^5
\left(O^{\rm ncw}_{L jk} O^{\rm cnw}_{L ki} +
O^{\rm ncw}_{R jk} O^{\rm cnw}_{R ki}\right) \bigg\{
2 B_1(p^2,m^2_k,m^2_W) + B_0(p^2,m^2_k,m^2_W) 
\nonumber\\&&
-\xi B_0(p^2,m^2_k,\xi m^2_W)
- \frac{m_k^2-p^2}{m^2_W} \Big[ B_1(p^2,m_W^2,m_k^2)
- B_1(p^2,\xi m_W^2,m_k^2)
\Big]\bigg\}
\label{Wgraph}\\ \vb{24}
(\Pi^V_{ij})^W &\hskip -5pt=\hskip -5pt& \frac{1}{16\pi^2}\,
\sum_{k=1}^5 \left(O^{\rm ncw}_{L jk} O^{\rm cnw}_{R ki} + O^{\rm
    ncw}_{R jk} O^{\rm cnw}_{L ki}\right)m_k \Big[ 3
B_0(p^2,m^2_k,m^2_W) + \xi B_0(p^2,m^2_k,\xi m^2_W) \Big] \nonumber
\eeqa This graph introduces an explicit dependence on the gauge
parameter $\xi$.  The other self energy graph with $\xi$ dependence is
the one that includes the charged Goldstone boson. The charged
Goldstone boson is one of the eight charged scalars $S^+_k$ resulting
from mixing between the two charged Higgs fields and the six charged
sleptons.  This  contribution is

\begin{center}
\begin{picture}(120,40)(0,22) % y_2 for equation position
\ArrowLine(5,25)(40,25)
\ArrowArc(60,25)(20,180,0)
\DashCArc(60,25)(20,0,180){4}
\ArrowLine(80,25)(115,25)
\Text(25,34)[]{$F^0_i$}
\Text(100,34)[]{$F^0_j$}
\Text(75,50)[]{$S^+_r$}
\Text(75,0)[]{$F^+_k$}
\end{picture}
$
=i\,(\slash{p}\,{\bf\Sigma}_{ij}^V-\Pi^V_{ij})^{S^+}+...
$
\end{center}
\vspace{5pt} \hfill \\
\noindent
where again, the dots indicate terms proportional to $\gamma_5$, and
\beqa
(\Sigma^V_{ij})^{S^+}&=& -\frac{1}{16\pi^2}\, \sum_{r=1}^8\, \sum_{k=1}^5
 \left(O^{\rm ncs}_{R jkr} O^{\rm cns}_{L kir} +
O^{\rm ncs}_{L jkr} O^{\rm cns}_{R kir}\right)\, B_1(p^2,m^2_k,m^2_r)\cr
\vb{24}
(\Pi^V_{ij})^{S^+}&=& -\frac{1}{16\pi^2}\, \sum_{r=1}^8\, \sum_{k=1}^5
\left(O^{\rm ncs}_{L jkr} O^{\rm cns}_{L kir} +
O^{\rm ncs}_{R jkr} O^{\rm cns}_{R kir}\right)\, m_k\, B_0(p^2,m^2_k,m^2_r)
\label{chascaloop}
\eeqa
with the couplings given in Appendix B. Nevertheless, gauge dependence
is not canceled after combining eqs.~(\ref{Wgraph}) and (\ref{chascaloop}).
In order to achieve it the inclusion of the Goldstone boson tadpole 
graphs, left over from the tadpole equations, is necessary:

\begin{center}
\begin{picture}(80,60)(0,26) % y_2 for equation position
\DashCArc(20,50)(20,0,360){4}
\DashLine(20,30)(20,0){4}
\ArrowLine(-20,0)(20,0)
\ArrowLine(20,0)(60,0)
\Text(45,70)[]{$G^{\pm}$}
\Text(30,15)[]{$S^0_k$}
\Text(-20,8)[]{$F^0_i$}
\Text(65,8)[]{$F^0_j$}
\end{picture}
$
=i\,(\slash{p}\,{\bf\Sigma}_{ij}^V-\Pi^V_{ij})^{Tad}+...
$
\end{center}
\vspace{5pt} \hfill \\
\noindent
where $({\bf\Sigma}_{ij}^V)^{Tad}=0$ and
\begin{eqnarray}
(\Pi^V_{ij})^{Tad}&=& -\frac{1}{32\pi^2}\, \sum_{k=1}^5
\left(O^{\rm nns}_{L jik} + O^{\rm nns}_{R jik} \right)
\frac{1}{m_{S^0_k}} g^{S^0S^{+}S^{-}}_{k\,G^+G^-}
A_0(\xi m^2_W) 
\end{eqnarray}
The $A_0$, $B_0$ and $B_1$ appearing above are Passarino-Veltman
functions \cite{Passarino:1979jh}, $g^{S^0S^{+}S^{-}}_{k\,G^+G^-}$
being the neutral scalar coupling to a pair of charged Goldstone
bosons, and $O^{\rm nns}$ the neutral scalar couplings to a pair of
neutral fermions (neutralino/neutrinos).
Numerically, we have checked that, by adding the Goldstone tadpoles to
the self energies our results do not change by varying the gauge
parameter from $\xi=1$ to $\xi=10^9$, thus establishing the gauge
invariance of the calculation.
Similarly we have also checked that the corresponding set of diagrams
involving the neutral gauge boson tadpole ($Z$) plus neutral ghost
tadpoles is gauge invariant and, similarly, the contribution to the
self energies due to $Z$ exchange plus neutral pseudoscalars and
neutral Goldstones is also gauge invariant.

Before we close this section we would like to add a short discussion
on the basic structure of the loops which will be useful in the
following. It is useful to do this in the approximation where the \rp
parameters are small, as discussed above.  As seen from the expression
for $ m_{eff}$, at tree-level the effective neutrino mass matrix in
this limit has the structure $m_{ij} \sim \Lambda_i \Lambda_j$, and at
this level of sophistication neutrino angles are simple functions of
ratios of $\Lambda_i/\Lambda_j$. The one-loop corrections, however, in
general destroy this simple picture.  This can be seen as follows.
The one-loop corrections have the general form,
\beq
(\Sigma_{ij},\Pi_{ij}) \sim \sum ({\cal O}_{i,a} {\cal O}_{j,b} 
            + {\cal O}_{i,c}{\cal O}_{j,d}) (B_1,m B_0)
\label{gen1lp}
\eeq
where the ${\cal O}$ stand symbolically for the various couplings. 
Now, since the expansion matrix $\xi$, defined in \eq{defxi} can 
be written as $\xi_{i\alpha} \sim f_{\alpha} \epsilon_i +
g_{\alpha} \Lambda_i$ (see \eq{xielementos}) a product of two 
couplings involving neutrino-neutralino mixing has the general 
structure,

\beq
{\cal O}_{i,a} {\cal O}_{j,b} \sim (f_{\alpha} \epsilon_i +
g_{\alpha} \Lambda_i) \times (f'_{\alpha} \epsilon_j +
g'_{\alpha} \Lambda_j) \times F(...)
\label{genstruc}
\eeq
where all the other dependence on the SUSY parameters has been 
hidden symbolically in $F(...)$. The one-loop corrections therefore 
also carry a certain index structure, which can be written as 
\beq
m^{1-loop}_{ij} \sim a \epsilon_i \epsilon_j +
                     b (\epsilon_i \Lambda_j + \Lambda_i \epsilon_j ) +
                     c \Lambda_i \Lambda_j
\label{symb1lp}
\eeq 
where $a,b,c$ are again complicated functions of SUSY parameters
involving couplings, the Passarino-Veltman functions etc. Clearly, the
terms proportional to $c$ in \eq{symb1lp} above will lead only to a
renormalization of the heaviest neutrino mass eigenstate.  On the
other hand the terms proportional to $a$ in \eq{symb1lp} are genuine
loop corrections.  Consider the simple case where all $ \Lambda_i
\equiv 0$.  Clearly in this case the tree-level neutrino mass is
absent, but the one-loop effective neutrino mass has the same index
structure as before, but now in terms of $\epsilon_{i,j}$'s instead of
$\Lambda_{i,j}$'s.  In this idealized case angles are given as simple
funtions of $\epsilon_i$ ratios.
For non-zero $\Lambda_i$ the terms proportional to $b$ in eq.
\eq{symb1lp}, however, destroy this simple picture. Any mismatch
between $\epsilon_i/\epsilon_j$ and $\Lambda_i/\Lambda_j$ will lead,
in general, to a very complex parameter dependence of the neutrino
angles.

\section{Numerical Results on Neutrino Masses and Mixings}
\label{OneLoopMasses}

Here we collect our numerical results on neutrino masses and mixings.
As we have seen, a characteristic of the BRpV model is the appearence
of vacuum expectation values for the sneutrino fields, $v_i'$ which
imply a tree--level mass for one of the neutrinos given by
\eq{mnutree}.  The one--loop--corrected neutrino mass matrix gives
important contributions to the heaviest neutrino mass which we have
determined through the renormalization procedure sketched above.

\begin{figure}
\setlength{\unitlength}{1mm}
\begin{picture}(50,70)
\put(-37,-150)
{\mbox{\epsfig{figure=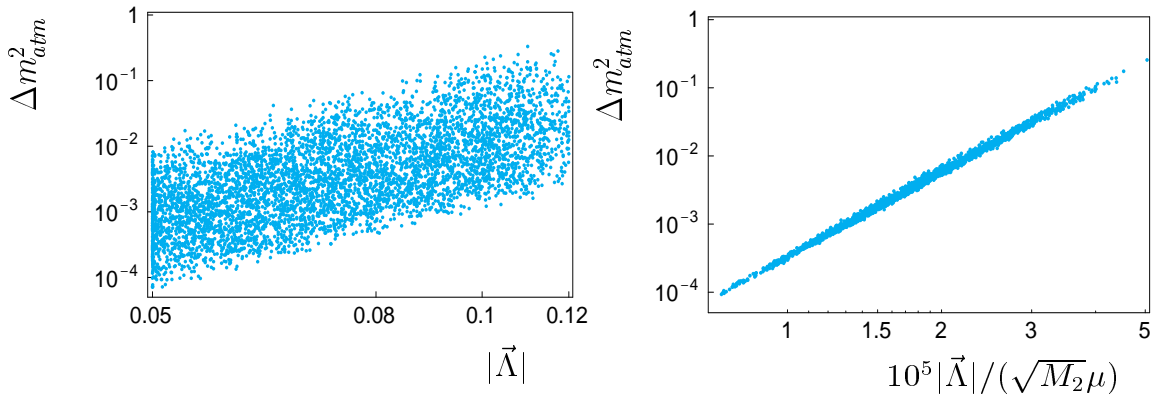,height=28.0cm,width=24.0cm}}}
\end{picture}
\caption[]{Example of calculated $\Delta m^2_{atm}$ as a function of 
  (left) the alignment parameter ${\vec \Lambda}$ and (right), as
  function of $|{\vec \Lambda}|/(\sqrt{M_2} \mu)$, all of these
  expressed in GeV. The figure shows that \eq{mnutree} can be used to
  fix the relative size of R-parity breaking parameters to obtain the
  correct $\Delta m^2_{atm}$.}
\label{atmmass}
\end{figure}

\begin{figure}
\setlength{\unitlength}{1mm}
\begin{picture}(50,70)
\put(-37,-150)
{\mbox{\epsfig{figure=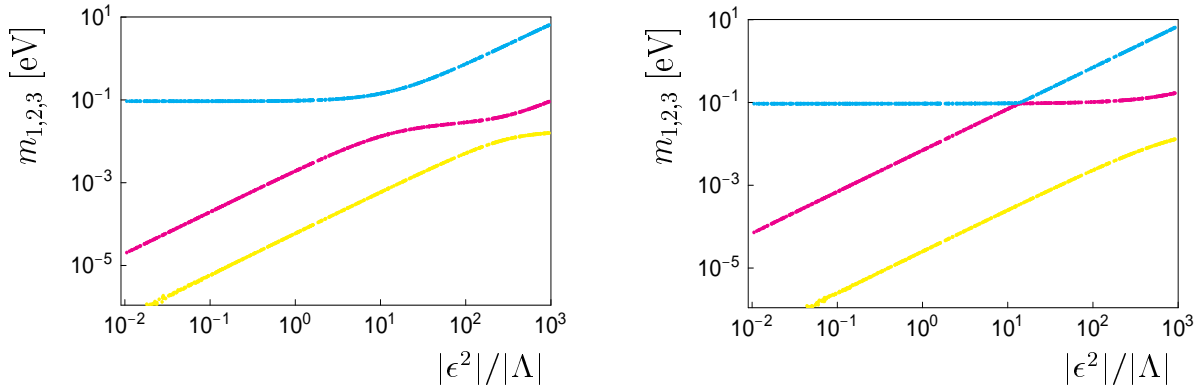,height=28.0cm,width=24.0cm}}}
\end{picture}
\caption[]{Example of calculated neutrino masses in units of eV as a 
  function of $|\epsilon^2|/|\Lambda|$, for a particular though
  typical choice for the other parameters (see text), illustrating the
  relative importance of tree versus loop-induced neutrino masses.}
\label{mivepsolam}
\end{figure}

First we note that in order to solve the atmospheric $and$ solar
neutrino problem one requires $m^{1-loop} \ll m^{tree}$. If this is
fullfilled it is essentially trivial within our model to solve the
atmospheric neutrino problem. It is simply equivalent to choosing an
adequate size of the alignment vector $|{\vec \Lambda}|^2$, as can be
seen from \eq{mnutree} and is also demonstrated in \fig{atmmass}.
However there are regions of parameters where the one-loop
contributions are comparable to the tree-level neutrino mass. This is
discussed in quite some detail below, where we give an illustrative
parameter study in order to isolate the main features of the
dependence on the underlying parameters. First we get a rough idea of
the magnitude of the neutrino masses including the one-loop
corrections by displaying in \fig{mivepsolam} the three lightest
eigenvalues of the neutrino/neutralino mass matrix as a function of
the parameter $|\epsilon^2|/|\Lambda|$.  Other parameters are fixed as
follows:
a) MSSM parameters: $m_0 = \mu = 500$ GeV, $M_2 = 200$ GeV, 
$\tan\beta = 5$, $B = -A = m_0$. 
b) RPV parameters: $|\Lambda| = 0.16$ $GeV^2$, $10 \Lambda_e =
\Lambda_{\mu} = \Lambda_{\tau}$ and
$\epsilon_1=\epsilon_2=\epsilon_3$. In the left panel we give the
predicted masses in the general case, while in the one on the right we
apply the sign condition, 
\beq
(\epsilon_{\mu}/\epsilon_{\tau})\times 
(\Lambda_{\mu}/\Lambda_{\tau}) \le 0
\label{signcondition}
\eeq
to be discussed in more detail below. 


One notices that the parameter $|\epsilon^2|/|\Lambda|$ determines the
importance of the loop contribution relative to the tree-level-induced
masses. For example, from the right panel one sees that, below
$|\epsilon^2|/|\Lambda| \ll 10$ the heaviest neutrino mass $m_3$ is
mainly a tree-level mass, while for $|\epsilon^2|/|\Lambda| \gsim 10$
the loop--induced masses are important relative to the tree--level
one.  Similar results are obtained for other choices of MSSM
parameters.

\begin{figure}
\setlength{\unitlength}{1mm}
\begin{picture}(50,70)
\put(-37,-150)
{\mbox{\epsfig{figure=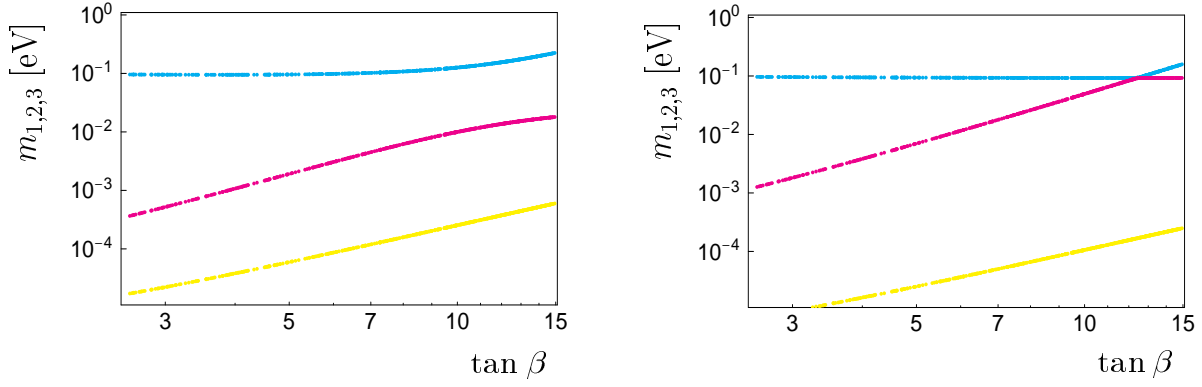,height=28.0cm,width=24.0cm}}}
\end{picture}
\caption[]{The neutrino mass spectrum versus $\tan\beta$, for parameters 
otherwise chosen as in \fig{mivepsolam}. The importance of loops 
increases strongly with $\tan\beta$. }
\label{mivtbeta}
\end{figure}

\begin{figure}
\setlength{\unitlength}{1mm}
\begin{picture}(50,70)
\put(-37,-150)
{\mbox{\epsfig{figure=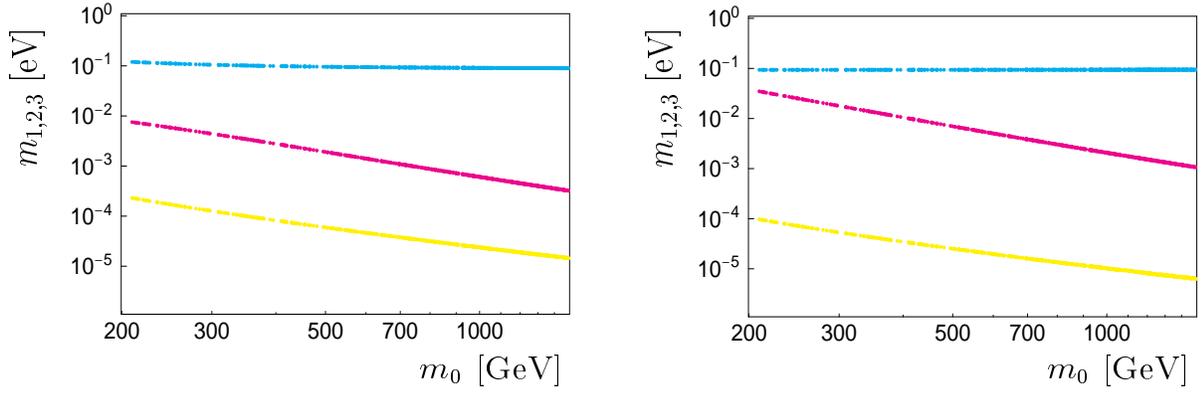,height=28.0cm,width=24.0cm}}}
\end{picture}
\caption[]{The neutrino mass spectrum versus $m_0$, for parameters 
otherwise chosen as in \fig{mivepsolam}. The importance of loops 
decreases with increasing $m_0$. }
\label{mivm0}
\end{figure}

It is also interesting to analyse the dependence of the neutrino mass
spectrum obtained in this model as a function of other supersymmetric
parameters. In \fig{mivtbeta} we show the three lightest eigenvalues
of the neutrino neutralino mass matrix as a function of $\tan\beta$,
keeping the other parameters fixed as in \fig{mivepsolam}, fixing
$\epsilon^2/|\Lambda| =1$. Again in the right figure we have applied
the sign condition discussed in more details below. Loop contributions are 
very strongly correlated with $\tan\beta$.
Similarly one can compute the three lightest eigenvalues of the
neutrino/neutralino mass matrix as a function of $m_0$, as shown in
\fig{mivm0}. Larger $m_0$ leads to smaller loop masses, as expected.
From \fig{mivepsolam}-\fig{mivm0} one sees that, as expected,
the pattern of neutrino masses obtained in the bilinear \rp scenario,
for almost all choices of parameters, is a \emph{hierarchical} one.

In the above we have not paid attention to whether or not the
parameter values used in the evaluation of the neutrino mass spectrum
are indeed solutions of the minimization tadpole conditions of the
Higgs potential.  We now move to a more careful study of the magnitude
of the neutrino mass spectrum derived in the \rp scenario.

In order to proceed further with the discussion of the solutions to
the solar neutrino anomalies in this model we must distinguish two
cases:
\begin{enumerate}
\item unified universal boundary conditions on the soft SUSY breaking
  terms   (SUGRA case, for short)
\item non--universal boundary conditions on the soft SUSY breaking
  terms (MSSM case, for short)
\end{enumerate}
In what follows we refer to these two possibilities as SUGRA and MSSM
cases, accordingly. 

\begin{figure}
\setlength{\unitlength}{1mm}
\begin{picture}(50,70)
\put(-37,-150)
{\mbox{\epsfig{figure=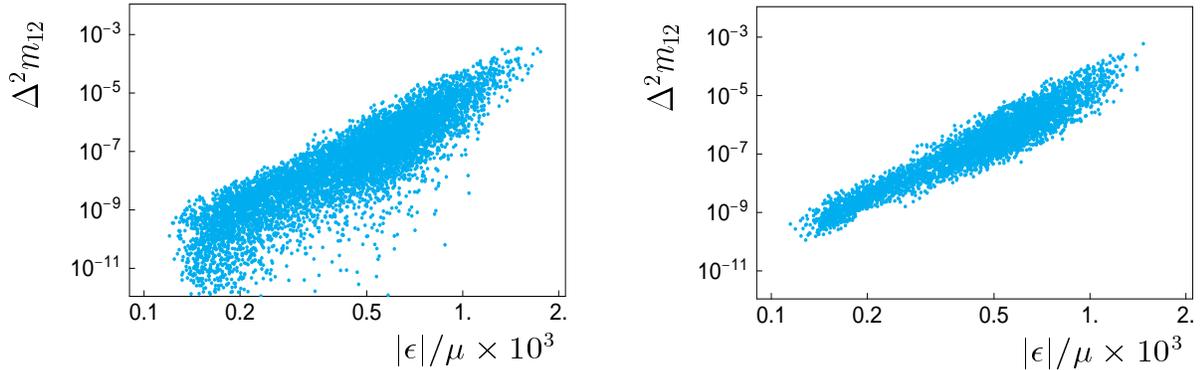,height=28.0cm,width=24.0cm}}}
\end{picture}
\caption[]{$\Delta m^2_{sol}$ versus $|\epsilon|/\mu$ for $\mu \le 0$ (left)
  and $\mu > 0$ (right).}
\label{delm12vmu}
\end{figure}

For the analysis of the neutrino {\em masses} these two scenarios are
very similar so we focus on the case where the low--scale paramaters
are derivable from a universal supergravity scheme.
In \fig{delm12vmu} we show the mass squared difference $\Delta
m^2_{12}$ which is relevant for the analysis of neutrino oscillations
and therefore relevant to the interpretation of solar data, as a
function of the parameter $|\epsilon|/\mu$. In the left panel we
display $\mu \le 0$ while on the right panel $\mu > 0$.  Small values
prefer $\Delta m^2_{12}$ in the range of the vacuum solution to the
solar neutrino problem, while large values give masses in the range of
the MSW solutions.

Points shown in the following figures were obtained scanning the
relevant parameters randomly over the region: $M_2$ and $|\mu|$ from 0
to 500 GeV, $m_0$ [0.2 TeV, 1.0 TeV], $a_0$ and $b_0$ [-3,3] and
$\tan\beta$ [2.5,10], and for the \rp parameters,
$|\Lambda_\mu/\Lambda_\tau|= 0.8-1.25$, $\epsilon_\mu/\epsilon_\tau =
0.8-1.25$, $|\Lambda_e/\Lambda_\tau|= 0.05-0.1$,
$\epsilon_e/\epsilon_\tau = 0.6-1.25$ and $|\Lambda| = 0.05-0.12$
$GeV^2$. They were subsequently tested for consistency with the
minimization (tadpole) conditions of the Higgs potential and for
phenomenological constraints from supersymmetric particle searches.

\begin{figure}
\setlength{\unitlength}{1mm}
\begin{picture}(50,70)
\put(-30,-100)
{\mbox{\epsfig{figure=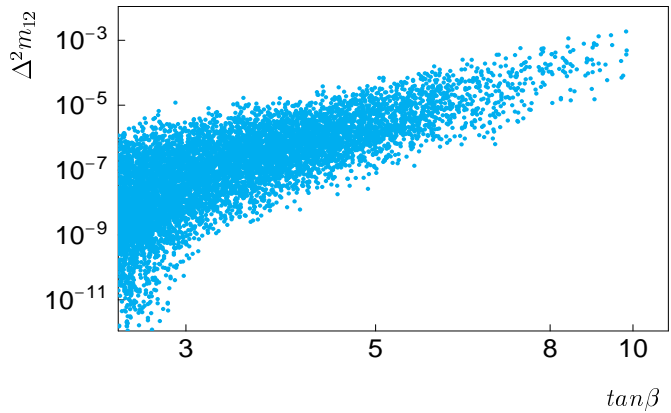,height=22.0cm,width=18.0cm}}}
\end{picture}
\caption[]{$\Delta m^2_{12}$ versus $\tan\beta$ for points which 
  solve the atmospheric neutrino problem.}
\label{delm12vtb}
\end{figure}

One can also explicitly determine the attainable range of
$\Delta m^2_{12}$ for which the corresponding $\Delta m^2_{23}$ (see
below) lies in the range required for the correct interpretation of
the atmospheric neutrino data. The result obtained is displayed in
\fig{delm12vtb} in which we show $\Delta m^2_{12}$ as function of 
$tan\beta$ for those points which solve the atmospheric neutrino problem.

We now turn to the discussion of the three neutrino mixing angles and
of how they must be identified in terms our our underlying
parameters.  Following the usual convention the relation
 \beq
 {\cal N}^{1L} = {\cal N'}\ {\cal N}
 \eeq
\beq
\nu_{\alpha} = U_{\alpha k}\,  \nu_k
\eeq
connecting mass-eigenstate and weak-eigenstate neutrinos are recovered
in our notation as
\beq
U_{\alpha k}= {\cal N}^{1L}_{4+k, 4+\alpha} 
\eeq
where the mixing coefficients \N are determined numerically by
diagonalizing the neutral fermion mass matrix. 
Note that, without loss of generality, in the bilinear model one can
always choose as basis the one in which the charged lepton mass matrix
is already diagonal.
The neutrino mixing angles relevant in the interpretation of
solar and atmospheric data are identified as (if $U_{e3} \ll 1$, 
as indicated by the atmospheric data and the reactor neutrino 
constraints). 
\beq
\sin^2 (2 \theta_{13}) = 4\, U_{\mu 3}^2 (1 -U_{\mu 3}^2)
\eeq
\beq
\sin^2 (2 \theta_{12}) = 4\, U_{e1}^2 U_{e2}^2
\eeq
\begin{figure}
\setlength{\unitlength}{1mm}
\begin{picture}(50,70)
\put(-30,-100)
{\mbox{\epsfig{figure=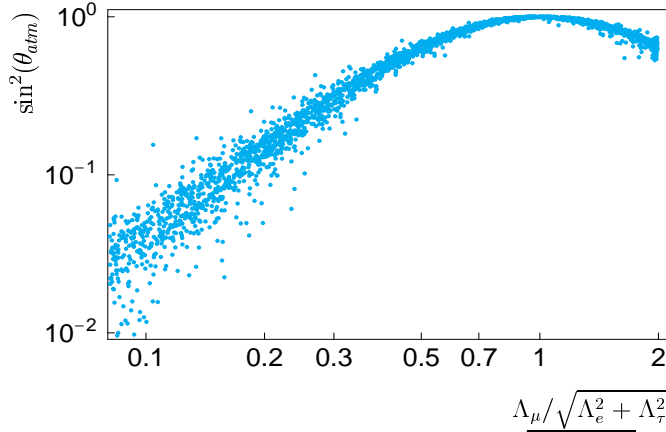,height=22.0cm,width=18.0cm}}}
\end{picture}
\caption{Atmospheric angle versus 
$\Lambda_{\mu}/\sqrt{\Lambda_{e}^2+\Lambda_{\tau}^2}$. Maximality 
is obtained for $\Lambda_{\mu} \simeq \Lambda_{\tau}$ if 
$\Lambda_{e}$ is smaller than the other two (see \fig{setchoozcnstr}). }
\label{setatmangle}
\end{figure}
Note that the maximality of the atmospheric angle is achieved for
$\Lambda_\mu = \Lambda_\tau$ (see \fig{setatmangle}) and $\Lambda_e$
is smaller than the other two, as required by the Chooz data (see
below).
In fact we have found~\cite{numass}, that if $\epsilon^2 /\Lambda \ll
10 $ then the approximate formula holds

\beq
U_{\alpha 3} \approx \Lambda_{\alpha}/|{\vec \Lambda}|
\eeq

\begin{figure}
\setlength{\unitlength}{1mm}
\begin{picture}(50,70)
\put(-30,-100)
{\mbox{\epsfig{figure=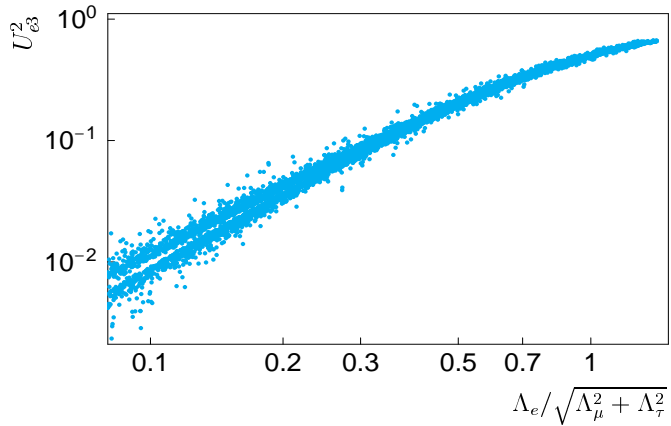,height=22.0cm,width=18.0cm}}}
\end{picture}
\caption{$U_{e3}^2$ versus
 $\Lambda_{e}/\sqrt{\Lambda_{\mu}^2+\Lambda_{\tau}^2}$. To obey the 
experimental bound $U_{e3}^2 \le 0.05$ $\Lambda_{e}$ must be smaller 
than $\Lambda_{\mu},\Lambda_{\tau}$. }
\label{setchoozcnstr}
\end{figure}

In \fig{setchoozcnstr} we show the expected magnutide of $U_{e3}^2$
versus the relevant ratio of \rp parameters. In order to comply with
the reactor data from the Chooz experiment one should have $U_{e3}^2$
below 0.05. This implies a bound on $\Lambda_e$ which can be read off
from the figure.

The discussion on the solar mixing angle is more involved. First note
that it has no meaning before adding the one--loop corrections to the
neutrino mass, since in that limit the two low-lying neutrinos would
be degenerate in mass. 

In order to proceed further with the discussion of the solutions to
the solar neutrino problem in this model we must analyse carefully the
implications of \eq{vpiapprox}. Here it is important to distinguish
between case 1 (SUGRA) and case 2 (MSSM) discussed above.

\begin{figure}
\setlength{\unitlength}{1mm}
\begin{picture}(50,70)
\put(-30,-100)
{\mbox{\epsfig{figure=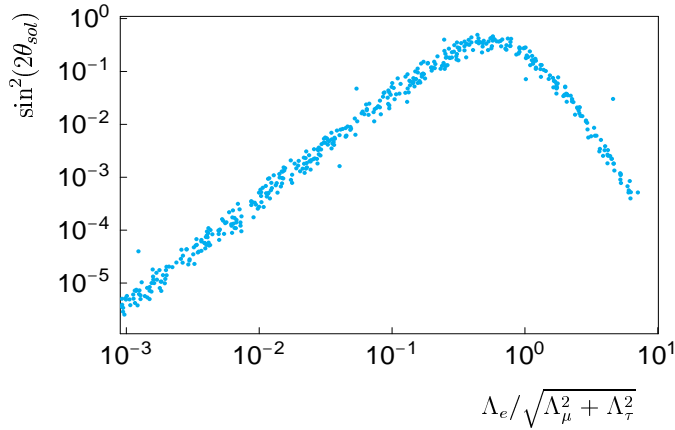,height=22.0cm,width=18.0cm}}}
\end{picture}
\vglue1cm
\caption{$sin^2(2\theta_{sol})$ versus
  $\Lambda_{e}/\sqrt{\Lambda_{\mu}^2+\Lambda_{\tau}^2}$ for the 
SUGRA case, for a discussion see text. }
\label{s2thsol002}
\end{figure}

In the SUGRA case by taking the ratio of the first two equations in
\eq{vpiapprox} 
\begin{equation}
\frac{\epsilon_e}{\epsilon_\mu} 
\frac{\Delta m^2_e-\tan\beta \mu \Delta B_e} 
     {\Delta m^2_\mu-\tan\beta \mu \Delta B_\mu}
\simeq \frac{\Lambda_e}{\Lambda_\mu} 
\label{ratio}
\end{equation}
we conclude that, since $\Lambda_e < \Lambda_\mu$ and, since the
relevant ratio of SUSY Soft--breaking terms is close to one, it
follows that $\sin^2 (2 \theta_{\odot})$ is small.  The predictions
for the solar angle as a function of the \rp breaking parameters is
indicated in \fig{s2thsol002}.

More precisely, the interpretation of the solar data \cite{MSW99} in
terms of the small angle MSW solution indicates that
\beq
\sin^2 (2 \theta_{\odot}) \lsim 10^{-3} - 10^{-2}
\eeq
and this in turn selects the required ratio of $\Lambda_e$ to
$\Lambda_\mu$ and $\Lambda_\tau$.  Therefore in this case the large
angle solutions, including the vacuum or just-so solutions do not fit
in the scheme.

\begin{figure}
\setlength{\unitlength}{1mm}
\begin{picture}(50,110)
\put(0,-70)
{\mbox{\epsfig{figure=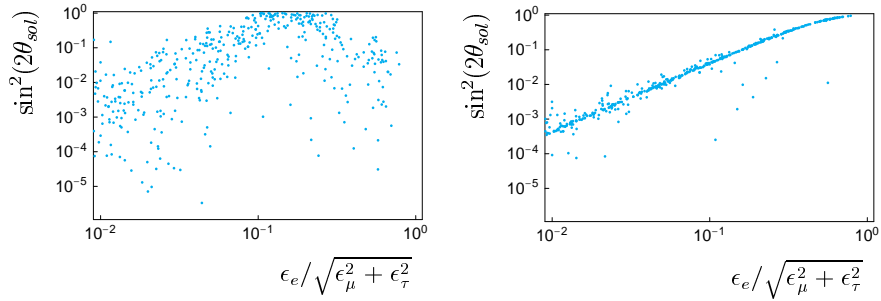,height=22.0cm,width=18.0cm}}}
\end{picture}
\vglue-5cm
\caption{$sin^2(2\theta_{sol})$ versus
  $\epsilon_{e}/\sqrt{\epsilon_{\mu}^2+\epsilon_{\tau}^2}$. The left
  panel corresponds to the case without the sign condition and the the
  right panel assumes the sign condition. }
\label{solsgn}
\end{figure}

We now move to the general MSSM case. In this case the ratio of SUSY
soft-breaking terms appearing in \eq{ratio} is in general arbitrary
and thus the ratios of $\Lambda_i/\Lambda_j$ is no longer tied up to
the ratios of $\epsilon_i/\epsilon_j$'s. This opens up the possibility
for large angle solutions to the solar neutrino problem. At first
sight it would seem that all predictivity of the solar angle is lost
in this case, as seen in left panel of \fig{solsgn}.

%
% Start Figure: Sign versus NO-Sign
%

%
% end of Figure: Sign versus NO-Sign
%

The ability of our model to determine the solar neutrino angle may be
understood in terms of \eq{symb1lp}. For example in the SUGRA case we
see from \eq{ratio} that the $\epsilon$ and $\Lambda$ ratios are fixed
within a narrow range, leading to the small mixing angle prediction
for the solution to the solar neutrino problems. There is however
another way to obtain predictivity for the general MSSM case, namely
by applying \eq{signcondition}.

The possibility of our model predicting the solar angle even in the
general MSSM case by assuming \eq{signcondition} can be understood as
follows.  Consider first the simplified limit $\Lambda_e \equiv 0$. In
this case $\nu_1 \equiv \nu_e$ at tree-level and there is no mixing at
all between the electron neutrino and the other two states, but a
finite mixing exists at one-loop, due to the terms proportional to
$\epsilon_e$.  In this case the sign condition, defined in
\eq{signcondition} introduces two more zeros into the matrix
proportional to $b$ in \eq{symb1lp} above, if $|\epsilon_{\mu}| \equiv
|\epsilon_{\tau}|$ and $|\Lambda_{\mu}| \equiv |\Lambda_{\tau}|$. This
fact simplifies the calculation of the solar angle very much, since
one of the neutrino eigenvectors (the one for $\nu_e$) has no
dependence on the $\Lambda_i$ ratios but only on the $\epsilon_i$
ratios. For a non-zero $\Lambda_e$ (and small departures from equality
of $\epsilon_{\mu},\epsilon_{\tau}$ or $|\Lambda_{\mu}|,
|\Lambda_{\tau}|$) this feature is destroyed and a $\Lambda_e$
dependence reintroduced in the solar angle. However, as long as the
one-loop contributions are smaller than the tree-level one and as long
as $\Lambda_e \ll \Lambda_{\mu,\tau}$, the ``cross-talk'' between the
$\Lambda_e$ and $\epsilon_e$ pieces is sufficiently small, such that
some predictivity of the solar angle is retained, as illustrated in
figure \fig{solsgn} (right panel).

The discussion on mixing angles may be summarized as follows.
In the case that one-loop corrections are not larger than the 
tree-level contributions, the approximate formula
\beq
U_{\alpha 3} \approx 
 \Lambda_\alpha / | \Lambda | 
\eeq
holds. This allows one to fix the atmospheric angle and at the same
time obey the CHOOZ constraint. For the solar angle, however, the
results depend on whether one wants to work in a SUGRA motivated
scenario or not. For the SUGRA scenario we have found that our model
allows only the small mixing angle MSW solution (SMA), while for the
general case also LMA and vacuum oscilation solutions are possible.

\section{Conclusions}

We have shown that the simplest unified extension of the Minimal
Supersymmetric Standard Model with bi-linear R--Parity violation
typically predicts a hierarchical neutrino mass spectrum, offering a
natural theory for the solar and atmospheric neutrino anomalies.  In
this model only one neutrino acquires mass due to mixing with
neutralinos, while the other two get mass only as a result of
radiative corrections.  We have performed a full one-loop calculation
of the effective neutrino mass matrix in the bi-linear \rp MSSM,
taking special care to achieve a manifestly gauge invariant
calculation and performing the renormalization of the heaviest
neutrino, needed in order to get reliable results.  The atmospheric
mass scale and maximal mixing angle arise from tree-level physics,
while the solar neutrino scale and oscillations follow from calculable
one-loop corrections. 

Under the assumption of universal boundary conditions for the
soft-supersymmetry breaking terms at the unification scale we find
that the atmospheric scale is calculable by the renormalization group
evolution due to the non-zero bottom quark Yukawa coupling. In this
case one predicts the small mixing angle (SMA) MSW solution to be the
only viable solution to the solar neutrino problem. 

In contrast, for the general MSSM model, where the above assumptions
are relaxed, one can implement a \emph{bi-maximal} \cite{bim} neutrino
mixing scheme, in which the solar neutrino problem is accounted for
through large mixing angle solutions, either MSW or just-so. A great
advantage of our approach is that the parameters required in order to
solve the neutrino anomalies can be independently tested at high
energy accelerators, as originally proposed in
\cite{ProjectiveMassMatrix}.  In fact, as shown in ref.
\cite{numass,Vissani} the bilinear \rp model predicts the lightest
supersymmetic particle (LSP) decay to be observable at high-energy
colliders, since the expected decay path can easily be shorter the
typical detector sizes.  This happens despite the smallness of
neutrino masses indicated by the SuperKamiokande data.  This provides
a way to test this solution of the atmospheric and solar neutrino
anomalies and potentially discriminate between the large and small
mixing solutions to the solar neutrino problem.

\section*{Acknowledgements}

This work was supported by DGICYT grant PB98-0693 and by the EEC under
the TMR contract ERBFMRX-CT96-0090.  M.H. has been supported by the
Marie-Curie program under grant No ERBFMBICT983000 and W.P. by a
fellowship from the Spanish Ministry of Culture under the contract
SB97-BU0475382. M.A.D. was supported by CONICYT grant No. 1000539.

\newpage

\appendix

\section{Mass Matrices} 
\label{MassMatrices}

\subsection{Scalar Mass Matrices}

\subsubsection{Charged Scalars}
 
The mass matrix of the charged scalar sector follows from the 
quadratic terms in the scalar potential 

\begin{equation} 
V_{quadratic}= S'^-
\bold{M_{S^{\pm}}^2} S'^+
\label{eq:Vquadratic} 
\end{equation} 
where the unrotated charged scalars are
$S'^+=(H_d^+, H_u^+, \widetilde{e}_L^+, \widetilde{\mu}_L^+
\widetilde{\tau}_L^+,\widetilde{e}_R^+, \widetilde{\mu}_R^+
\widetilde{\tau}_R^+)$. 
For convenience we will divide this $(8\times8)$ matrix into 
blocks in the following way: 
\begin{equation} 
\bold{M_{S^{\pm}}^2}=\left[\matrix{ 
{\bold M_{HH}^2} & {\bold M_{H\tilde\ell}^{2T}} \cr 
{\bold M_{H\tilde\ell}^2} & {\bold M_{\tilde\ell\tilde\ell}^2} 
}\right] + \xi m_W^2 
\left[\matrix{
{\bold M_{A}^2} & {\bold M_{B}^{2T}} \cr 
{\bold M_{B}^2} & {\bold M_{C}^2} 
}\right] 
\label{eq:subdivM} 
\end{equation} 
where the charged Higgs block is 
\begin{eqnarray} 
{\bold M_{HH}^2}&=& 
\left[\matrix{ 
B\mu{{v_u}\over{v_d}}+\quarter g^2(v_2^2-\sum_{i=1}^3 v_i^2)
+{{t_d}\over{v_d}} & B\mu+\quarter g^2 v_d v_u \cr
+\mu \sum_{i=1}^3 \epsilon_i {{v_i}\over{v_d}}
+\half \sum_{i,j=1}^3 v_i \left(h_E h_E^{\dagger}\right)_{ij} v_j 
& \cr
\vb{30} 
B\mu+\quarter g^2v_d v_u 
& B\mu{{v_d}\over{v_u}}+\quarter g^2(v_d^2+\sum_{i=1}^3 v_i^2)
\cr
&- \sum_{i=1}^3 B_i\epsilon_i {{v_i}\over{v_u}} +{{t_u}\over{v_u}}  
}\right] 
%\nonumber 
\label{eq:subMHH} 
\end{eqnarray} 
This matrix  reduces to the usual charged Higgs mass matrix in the 
MSSM when we set $v_i=\epsilon_i=0$ and we call $m_{12}^2=B\mu$. 
The slepton block is given by 
\begin{eqnarray} 
\label{eq:subtautau}
{\bold M_{\tilde\ell\tilde\ell}^2}&=& 
\left[\matrix{ {\bold M^2_{LL}}& {\bold M^2_{LR}}\cr
{\bold M^2_{RL}}& {\bold M^2_{RR}}
}\right]
\end{eqnarray} 
where
\beqa
\left({\bold M^2_{LL}}\right)_{ij}
&=&\half v_d^2 \left(h_E^* h_E^T\right)_{ij} 
+\quarter g^2\left(-\sum_{k=1}^3 v_k^2-v_d^2+v_u^2\right)\delta_{ij}
+\quarter g^2 v_i v_j -\frac{v_u}{v_i} B_i \epsilon_i \delta_{ij}
+\frac{t_i}{v_i} \delta_{ij}\cr
\vb{18}
&&+\mu \frac{v_d}{v_i}\epsilon_i \delta_{ij}
-\epsilon_i \left(\sum_{k=1}^3 \frac{v_k}{v_i} \epsilon_k\right) \delta_{ij}
+\epsilon_i \epsilon_j
+M^2_{L ji}\cr
\vb{18}
&&-\half \sum_{k=1}^3 \frac{v_k}{v_i} 
\left(M^2_{L ik}+M^2_{L ki}\right)\delta_{ij}
\eeqa
\beq
{\bold M^2_{LR}}
=\frac{1}{\sqrt{2}}\, \left(v_d A_E^* -\mu v_u h_E^* \right)
\eeq
\beq
{\bold M^2_{RL}}=\left({\bold M^2_{LR}}\right)^{\dagger}
\eeq
\begin{eqnarray} 
\left({\bold M_{RR}^2}\right)_{ij}&=&
\quarter g'^2\left(-\sum_{k=1}^3 v_k^2-v_d^2+v_u^2\right)\delta_{ij}
+\half\, v_d^2 \left( h_E^T h_E^*\right)_{ij} \cr
\vb{18}
&&+\left(\sum_{k=1}^3 \left(h_E^T\right)_{ik} v_k\right) 
\left(\sum_{s=1}^3 \left(h_E^*\right)_{sj} v_s\right) 
+M^2_{R ji}
\end{eqnarray} 
We recover the usual stau mass matrix again by replacing  
$v_i=\epsilon_i=0$ (note that we need to replace the expression of 
the tadpole $t_i$ in eq.~(\ref{eq:tadpoles}) before taking the limit). 
The mixing between the charged Higgs sector and the slepton sector is 
given by the following $6\times2$ block (repeated indices are not
summed unless an explicit sum appears): 
\begin{equation} 
{\bold M_{H\tilde\ell}^2}=
\left[\matrix{ 
\ds
-\mu\epsilon_i-\half v_d \sum_{k=1}^3 \left(h_E^*
h_E^T\right)_{ik} v_k +\quarter g^2v_d v_i 
& -B_i\epsilon_i+\quarter g^2v_uv_i 
\cr 
\vb{20}
\ds
-\sqrthalf v_u \sum_{k=1}^3\, 
\left(h_E^T\right)_{ik}\epsilon_k
-\sqrthalf \sum_{k=1}^3\, 
\left(A_E^T\right)_{ik}v_k
&\ds
 -\sqrthalf \sum_{k=1}^3
\left(h_E^T\right)_{ik}(\mu v_k+\epsilon_kv_d) 
}\right] 
\label{eq:subHtau} 
\end{equation} 
and as expected, this mixing vanishes in the limit $v_i=\epsilon_i=0$. 
The charged scalar mass matrix in eq.~(\ref{eq:subdivM}), 
after setting $t_u=t_d=t_i=0$, has determinant  equal to zero for 
$\xi=0$, since one of the eigenvectors corresponds to the charged 
Goldstone boson with zero eigenvalue. 

\ni
For our one loop calculations one has to had the gauge fixing. 
The part of the mass matrix in Eq.~(\ref{eq:subdivM}) that comes 
from the gauge fixing reads for the $(2\times2)$ $A$ block
\begin{equation} 
{\bold M_{A}^2}=
\left[\matrix{ 
\ds
\frac{v_d^2}{v^2}&\ds \frac{-v_uv_d}{v^2}\cr
\vb{18}
\ds
\frac{-v_uv_d}{v^2}&\ds \frac{v_u^2}{v^2}
}\right] 
\end{equation} 
for the $(6\times2)$ $B$ and the $(6\times6)$ $C$ blocks
\begin{equation} 
{\bold M_{B}^2}=
\left[\matrix{ 
\ds
\frac{v_iv_d}{v^2}&\ds \frac{-v_iv_u}{v^2}\cr
\vb{18}
0 & 0
}\right] 
\hskip 10mm ; \hskip 10mm
{\bold M_{C}^2}=
\left[\matrix{ 
{\bold M_{D}^2}&0\cr
\vb{18}
0 & 0
}\right] 
\end{equation} 
where the $(3\times 3)$ $D$ block is
\begin{equation} 
{\bold M_{D}^2}=
\left[\matrix{ 
\ds\frac{v_1^2}{v^2}&\ds \frac{v_1v_2}{v^2}&\ds \frac{v_1v_3}{v^2}\cr
\vb{18}
\ds\frac{v_2v_1}{v^2}&\ds \frac{v_2^2}{v^2}&\ds \frac{v_2v_3}{v^2}\cr
\vb{18}
\ds\frac{v_3v_1}{v^2}&\ds \frac{v_2v_3}{v^2}&\ds \frac{v_3^2}{v^2}
}\right] 
\end{equation} 
The charged scalar mass matrices are diagonalized by the following rotation
matrices,
\begin{equation} 
S_i^{\pm}=
{\bold R}^{S^{\pm}}_{ij} S^{\pm'}_j
\end{equation} 
with the eigenvalues  
$\rm{diag}(m^2_{S_1},\ldots,m^2_{S_8})=
{\bold R^{S^{\pm}}}\, {\bold M_{S^{\pm}}^2}\,
\left({\bold R^{S^{\pm}}}\right)^T$.

\subsubsection{CP--Even Neutral Scalars}

\ni
The quadratic scalar potential includes 
\begin{equation} 
V_{quadratic}=
\half[\sigma^0_d,\sigma^0_u,\tilde\nu_{i}^R] 
{\bold M^2_{S^0}}\left[\matrix{ 
\sigma^0_d \cr \sigma^0_u \cr \tilde\nu_{i}^R 
}\right]+ \cdots
\label{eq:NeutScalLag} 
\end{equation} 
where the 
neutral CP-even scalar sector mass matrix in eq.~(\ref{eq:NeutScalLag}) 
is given by 
\beq
{\bold M_{S^0}^2} =
\left[\matrix{
{\bold M_{SS}^2} &  {\bold M_{S\widetilde{\nu}_R}^2} \cr
\vb{16}
{\bold M_{S\widetilde{\nu}_R}^2}^T&
{\bold M_{\widetilde{\nu}_R \widetilde{\nu}_R}^2}  
}\right]
\eeq
where
\beq
{\bold M_{SS}^2}= 
\left[\matrix{ \ds
B\mu{{v_u}\over{v_d}}+\quarter g_Z^2v_d^2+\mu \sum_{k=1}^3\epsilon_k 
{{v_k}\over{v_1}}+{{t_d}\over{v_d}}  
& -B\mu-\quarter g^2_Zv_dv_u  
\cr 
\vb{18}
-B\mu-\quarter g^2_Zv_dv_u  
&\ds
 B\mu{{v_d}\over{v_u}}+\quarter g^2_Zv_u^2-\sum_{k=1}^3\, 
B_k\epsilon_k {{v_k}\over{v_2}}+{{t_u}\over{v_u}}  
}\right] 
\label{eq:MSS}
\eeq
\beq
{\bold M_{S\widetilde{\nu}_R}^2}= 
\left[\matrix{ 
-\mu\epsilon_i+\quarter g^2_Zv_dv_i  
\cr
\vb{18}
B_i\epsilon_i-\quarter g^2_Zv_uv_i  
}\right] \nonumber 
\eeq
and
\beqa
\left({\bold M_{\widetilde{\nu}_R \widetilde{\nu}_R}^2}\right)_{ij}
\hskip -10pt&=\hskip -10pt& 
\left(\mu\epsilon_i{{v_d}\over{v_i}}
- B_i\epsilon_i{{v_u}\over{v_i}}
-\epsilon_i \sum_{k=1}^3 \epsilon_k \frac{v_k}{v_i}
- \half \sum_{k=1}^3  \frac{v_k}{v_i} \left(M^2_{L ik}+M^2_{L ki}\right)
+ {{t_i}\over{v_i}}\right)\delta_{ij} +\quarter g^2_Zv_iv_j  \cr
\vb{20}
&&+ \epsilon_i \epsilon_j +\half \left(M^2_{L ij}+M^2_{L ji}\right)
\eeqa
where we have defined $g_Z^2\equiv g^2+g'^2$. In the upper--left 
$2\times2$ block, in the limit $v_i=\epsilon_i=0$, the reader can 
recognize the MSSM mass matrix corresponding to the CP--even neutral 
Higgs sector. 
To define the rotation matrices let us define the unrotated fields by
\beq
S'^0=(\sigma_d^0,\sigma_u^0,\widetilde{\nu}_1^R,
\widetilde{\nu}_2^R, \widetilde{\nu}_2^R) 
\eeq
Then the mass eigenstates are $S^0_i$ given by
\begin{equation} 
S^0_i=
{\bold R}^{S^0}_{ij} S'^0_j
\end{equation} 
with the eigenvalues  
$\rm{diag}(m^2_{S_1},\ldots,m^2_{S_5})=
{\bold R^{S^0}}\, {\bold M_{S^0}^2}\,
\left({\bold R}^{S^0}\right)^T$.

\subsubsection{CP--Odd Neutral Scalars}

\ni
The quadratic scalar potential includes 
\begin{equation} 
V_{quadratic}=\half[\varphi^0_1,\varphi^0_2,\tilde\nu_{i}^I] 
{\bold M^2_{P^0}}\left[\matrix{ 
\varphi^0_1 \cr \varphi^0_2 \cr \tilde\nu_{i}^I 
}\right] + \cdots
\label{eq:PseudoScalLag} 
\end{equation} 
where the CP-odd neutral scalar mass matrix is 
\beq
{\bold M_{P^0}^2} =
\left[\matrix{
{\bold M_{PP}^2} &  {\bold M_{P\widetilde{\nu}_I}^2} \cr
\vb{16}
{\bold M_{P\widetilde{\nu}_I}^2}^T&
{\bold M_{\widetilde{\nu}_I \widetilde{\nu}_I}^2}  
}\right]
+ \xi m_Z^2 
\left[\matrix{
{\bold M_{E}^2} & {\bold M_{F}^{2T}} \cr 
{\bold M_{F}^2} & {\bold M_{G}^2} 
}\right] 
\label{eq:MP0}
\eeq
where

\beq
{\bold M_{PP}^2}= 
\left[\matrix{ \ds
B\mu{{v_u}\over{v_d}}+\mu\sum_{k=1}^3
\epsilon_k{{v_k}\over{v_d}}+{{t_d}\over{v_d}} 
& B\mu \cr
\vb{20}
B\mu & \ds
B\mu{{v_d}\over{v_u}}-\sum_{k=1}^3
B_k\epsilon_k{{v_k}\over{v_u}}+{{t_u}\over{v_u}} 
}\right] 
\label{eq:MPP}
\eeq

\beq
{\bold M_{P\widetilde{\nu}_I}^2}= 
\left[\matrix{ 
-\mu\epsilon_i  
\cr
\vb{18}
-B_i\epsilon_i
}\right] \nonumber 
\eeq

\ni
and

\beqa
\left({\bold M_{\widetilde{\nu}_I \widetilde{\nu}_I}^2}\right)_{ij}&=& 
\left(\mu\epsilon_i{{v_d}\over{v_i}}
- B_i\epsilon_i{{v_u}\over{v_i}}
-\epsilon_i \sum_{k=1}^3 \epsilon_k \frac{v_k}{v_i}
- \half \sum_{k=1}^3  \frac{v_k}{v_i} \left(M^2_{L ik}+M^2_{L ki}\right)
+ {{t_i}\over{v_i}}\right)\delta_{ij}  \cr
\vb{20}
&&+ \epsilon_i \epsilon_j +\half \left(M^2_{L ij}+M^2_{L ji}\right)
\eeqa

Finally the part of the mass matrix in Eq.~(\ref{eq:MP0}) that comes 
from the gauge fixing reads for the $(2\times2)$ $E$ block
\begin{equation} 
{\bold M_{E}^2}=
\left[\matrix{ 
\ds
\frac{v_d^2}{v^2}&\ds \frac{-v_uv_d}{v^2}\cr
\vb{18}
\ds
\frac{-v_uv_d}{v^2}&\ds \frac{v_u^2}{v^2}
}\right] 
\end{equation} 
for the $(3\times2)$ $F$ block
\begin{equation} 
{\bold M_{F}^2}=
\left[\matrix{ 
\ds
\frac{v_iv_d}{v^2}&\ds \frac{-v_iv_u}{v^2}
}\right] 
\end{equation} 
and for the $(3\times3)$ $G$ block
\begin{equation} 
{\bold M_{G}^2}=
\left[\matrix{ 
\ds\frac{v_1^2}{v^2}&\ds \frac{v_1v_2}{v^2}&\ds \frac{v_1v_3}{v^2}\cr
\vb{18}
\ds\frac{v_2v_1}{v^2}&\ds \frac{v_2^2}{v^2}&\ds \frac{v_2v_3}{v^2}\cr
\vb{18}
\ds\frac{v_3v_1}{v^2}&\ds \frac{v_3v_2}{v^2}&\ds \frac{v_3^2}{v^2}\cr
}\right] 
\end{equation} 
The charged pseudo--scalar mass matrices are diagonalized by the 
following rotation matrices,
\begin{equation} 
P_i=
{\bold R}^{P^0}_{ij} P'_j
\end{equation} 
with the eigenvalues  
$\rm{diag}(m^2_{A_1},\ldots,m^2_{A_5})=
{\bold R}^{P^0}\, {\bold M_{P^0}^2}\,
\left({\bold R}^{P^0}\right)^T$.
where the unrotated fields are
\beq
P'^0=(\varphi_d^0,\varphi_u^0,\widetilde{\nu}_1^I,
\widetilde{\nu}_2^I, \widetilde{\nu}_2^I) 
\eeq

\subsubsection{Squark Mass Matrices}

In the unrotated basis $\widetilde{u}'_i=(\widetilde{u}_{Li},
\widetilde{u}_{Ri}^*)$ and
$\widetilde{d}'_i=(\widetilde{d}_{Li},
\widetilde{d}^*_{Ri})$ we get
\beq
V_{quadratic}=\half \widetilde{u}'^{\dagger}\, \bold{M_{\widetilde{u}}}^2\,
\widetilde{u}'
+\half \widetilde{d}'^{\dagger}\, \bold{M_{\widetilde{d}}}^2\,
\widetilde{d}'
\eeq
where
\beq
\bold{M_{\widetilde{q}}}^2=
\left(
\matrix{
\bold{M^2_{\widetilde{q}LL}}&\bold{M^2_{\widetilde{q}LR}}\cr
\vb{18}
\bold{M^2_{\widetilde{q}RL}}&\bold{M^2_{\widetilde{q}RR}}
}
\right)
\label{eq:MassSquarks}
\eeq
with $\widetilde{q}=(\widetilde{u},\widetilde{d})$. The blocks are
different for up and down type squarks. We have
\beqa
\bold{M^2_{\widetilde{u}LL}}&=&\smallfrac{1}{2}\, v_u^2\, h_U^* h_U^T + M^2_Q
+\smallfrac{1}{6}\, (4 m_W^2 -m_Z^2) \cos 2 \beta \cr
\vb{18}
\bold{M^2_{\widetilde{u}RR}}&=& \smallfrac{1}{2}\, v_u^2\, h_U^T h_U^* + M^2_U
+ \smallfrac{2}{3}( m_Z^2 -m_W^2) \cos 2 \beta \cr
\vb{18}
\bold{M^2_{\widetilde{u}LR}}&=&\frac{v_u}{\sqrt{2}}\, A_U^* -\mu
\frac{v_d}{\sqrt{2}}\, h_U^* +
\sum_{i=1}^3 \frac{ v_i}{\sqrt{2}}\, \epsilon_i\, h_U^* \cr
\vb{18}
\bold{M^2_{\widetilde{u}RL}}&=&\bold{M^2_{\widetilde{u}LR}}^{\dagger}
\eeqa
and
\beqa
\bold{M^2_{\widetilde{d}LL}}&=&\frac{1}{2}\, v_d^2\, h_D^* h_D^T + M^2_Q
-\smallfrac{1}{6}\, (2 m_W^2 + m_Z^2) \cos 2 \beta \cr
\vb{18}
\bold{M^2_{\widetilde{d}RR}}&=& \smallfrac{1}{2}\, v_d^2\, h_D^T h_D^* + M^2_D
- \smallfrac{1}{3}( m_Z^2 -m_W^2) \cos 2 \beta \cr
\vb{18}
\bold{M^2_{\widetilde{d}LR}}&=&\frac{v_d}{\sqrt{2}}\, A_D^* -\mu
\frac{v_u}{\sqrt{2}}\, h_D^* \cr
\vb{18}
\bold{M^2_{\widetilde{d}RL}}&=&\bold{M^2_{\widetilde{d}LR}}^{\dagger}
\eeqa
We define the mass eigenstates
\beq
\widetilde{q}=\bold{R^{\widetilde{q}}}\, \widetilde{q}'
\eeq
which implies
\beq
\widetilde{q}'_i=\bold{R^{\widetilde{q}}}_{ji}^{\ *} \, \widetilde{q}_j
\eeq
The rotation matrices are obtained from
\beq
\bold{R^{\widetilde{q}}}^{\ \dagger}
\left(\bold{M_{\widetilde{q}}^{diag}}\right)^2
\bold{R^{\widetilde{q}}}= \bold{M_{\widetilde{q}}}^2
\eeq
In our case the matrices in \Eq{eq:MassSquarks} are real and
therefore the rotation matrices $\bold{R^{\widetilde{q}}}$ are
orthogonal matrices.

\subsection{Chargino Mass Matrix}

The charginos mix with the charged leptons forming a set of  
five charged fermions $F_i^{\pm}$, $i=1,\ldots,5$ in two component spinor
notation. In a basis where  
$\psi^{+T}=(-i\lambda^+,\widetilde H_u^+,e_R^+,\mu_R^+,\tau_R^+)$ 
and $\psi^{-T}=(-i\lambda^-,\widetilde
H_d^-,e_L^-,\mu_L^-,\tau_L^-)$, the charged 
fermion mass terms in the Lagrangian are 
\begin{equation} 
{\cal L}_m=-{1\over 2}(\psi^{+T},\psi^{-T}) 
\left(\matrix{{\bold 0} & \bold M_C^T \cr {\bold M_C} &  
{\bold 0} }\right) 
\left(\matrix{\psi^+ \cr \psi^-}\right)+h.c. 
\label{eq:chFmterm} 
\end{equation} 
where the chargino/lepton mass matrix is given by

\begin{equation} 
{\bold M_C}=\left[\matrix{ 
M & {\textstyle{1\over{\sqrt{2}}}}gv_u & 0 &0& 0\cr 
\vb{18}
{\textstyle{1\over{\sqrt{2}}}}gv_d & \mu &  
-{\textstyle{1\over{\sqrt{2}}}}\left(h_E\right)_{11}v_1& 
-{\textstyle{1\over{\sqrt{2}}}}\left(h_E\right)_{22}v_2& 
-{\textstyle{1\over{\sqrt{2}}}}\left(h_E\right)_{33}v_3 \cr
\vb{18} 
{\textstyle{1\over{\sqrt{2}}}}gv_1 & -\epsilon_1 & 
{\textstyle{1\over{\sqrt{2}}}}\left(h_E\right)_{11}v_d&0&0\cr
{\textstyle{1\over{\sqrt{2}}}}gv_2 & -\epsilon_2 & 0&
{\textstyle{1\over{\sqrt{2}}}}\left(h_E\right)_{22}v_d&0\cr
{\textstyle{1\over{\sqrt{2}}}}gv_3 & -\epsilon_3 & 0&0&
{\textstyle{1\over{\sqrt{2}}}}\left(h_E\right)_{33}v_d}
\right] 
\label{eq:ChaM6x6} 
\end{equation} 
and $M$ is the $SU(2)$ gaugino soft mass. 
We note that chargino sector decouples from the lepton sector in the limit 
$\epsilon_i=v_i=0$. As in the MSSM, the chargino mass matrix is  
diagonalized by two rotation matrices $\bold U$ and $\bold V$ defined
by
\beq
F_i^-=U_{ij}\, \psi_j^- \quad ; \quad F_i^+=V_{ij}\, \psi_j^+
\eeq
Then 
\beq
\bold{U}^* \bold{M_C} \bold{V}^{-1} = \bold{M_{CD}}
\eeq
where $\bold{M_{CD}}$ is the diagonal charged fermion mass matrix.
To determine $\bold{U}$ and $\bold{V}$ we note that
\beq
\bold{M^2_{CD}}=\bold{V} \bold{M_C^{\dagger}} \bold{M_C} \bold{V}^{-1} =
\bold{U^*} \bold{M_C} \bold{M_C^{\dagger}} (\bold{U^*})^{-1}
\eeq
implying that $\bold{V}$ diagonalizes $\bold{M_C^{\dagger} M_C}$ and 
$\bold{U^*}$ diagonalizes $\bold{M_C M_C^{\dagger}}$. For future
reference we note that 
\beq
\psi^-_j=\bold{U^*}_{kj}\, F_k^- 
\quad ; \quad
\psi^+_j=\bold{V^*}_{kj}\, F_k^+ 
\eeq

\ni
In the previous expressions the $F_i^{\pm}$ are two component
spinors. We construct the four component Dirac spinors out of the two
component spinors with the
conventions\footnote{Here we depart from the conventions of
ref.~\cite{HabKaneGun} because we want the $e^-$, $\mu^-$ and $\tau^-$ 
to be the particles and not the anti--particles.},
\beq
\chi_i^-=\left(\matrix{
F_i^- \cr
\vb{18}
\overline{F_i^+}\cr}
\right)
\eeq

\section{The Couplings}
\label{couplings}
\subsection{The Neutralino Couplings}

Using four component spinor notation the relevant part of the
Lagrangian can be written as
 
\beqa
\hskip -15pt
{\cal L}&=&
\ovl{\chi_i^-}\, \gamma^{\mu}\,  \left ( O^{\rm cnw}_{L ij} P_L +
O^{\rm cnw}_{Rij} P_R \right) \chi_j^0  \  W_{\mu}^-  
+ \ovl{\chi_i^0}\, \gamma^{\mu}\,  \left( O^{\rm ncw}_{L ij} P_L +
O^{\rm ncw}_{Lij} P_R \right) \chi_j^- \ W_{\mu}^+ \cr
\vb{20}
&&+
\ovl{\chi_i^-}\, \,  \left ( O^{\rm cns}_{L ijk} P_L +
O^{\rm cns}_{Rijk} P_R \right) \chi_j^0  \  S_k^-  
+ \ovl{\chi_i^0}\, \,  \left( O^{\rm ncs}_{L jik} P_L +
O^{\rm ncs}_{Ljik} P_R \right) \chi_j^- \ S_k^+ \cr
\vb{20}
&&+\half  \ovl{\chi_i^0} \gamma^{\mu}\,
\left( O^{\rm nnz}_{Lij} P_L 
+ O^{\rm nnz}_{Rij} P_R \right) \chi_j^0 \ Z_{\mu}^0 
+\half  \ovl{\chi_i^0} \,
\left( O^{\rm nnh}_{L ijk} P_L 
+ O^{\rm nnh}_{R ijk} P_R \right) \chi_j^0 \ H_k^0 \cr
\vb{20}
&&+ i\, \half  \ovl{\chi_i^0} \,
\left( O^{\rm nna}_{L ijk} P_L 
+ O^{\rm nna}_{R ijk} P_R \right) \chi_j^0 \ A_k^0 \cr
\vb{20}
&& + \ovl{q_i}
\left( O^{\rm qns}_{Lijk} P_L 
+ O^{\rm qns}_{Rijk} P_R \right) \chi_j^0 \ {\tilde q}_{k} 
+ \ovl{\chi_i^0}
\left( O^{\rm nqs}_{Lijk} P_L 
+ O^{\rm nqs}_{Rijk} P_R \right) q_j  \ {\tilde q}_{k}^*
\eeqa  
where $q$ can be either $d$ or $u$. The various couplings are:

\subsubsection{Chargino--Neutralino--W}

\beq
\ba{l}\ds
O^{\rm cnw}_{L ij}= g\, \eta_i \eta_j
\left[- \bold{N^*}_{j2} \bold{U}_{i1} -\sqrthalf\,
\left( \bold{N^*}_{j3} \bold{U}_{i2} + 
\sum_{k=1}^3 \bold{N^*}_{j,4+k}\, \bold{U}_{i,2+k} \right) \right]\cr
\vb{22}
O^{\rm cnw}_{R ij}=g\, \left(- \bold{N}_{j2} \bold{V^*}_{i1} +\sqrthalf\,
\bold{N}_{j4} \bold{V^*}_{i2} \right)\cr
\vb{22}
O^{\rm ncw}_{L ij}= \left( O^{\rm cnw}_{L ji} \right)^*
\hskip 5mm ; \hskip 5mm
O^{\rm ncw}_{R ij}= \left( O^{\rm cnw}_{R ji} \right)^*
\ea 
\eeq

\subsubsection{Neutralino--Neutralino--Z}

\beq
\ba{l}\ds
O^{\rm nnz}_{L ij}= \frac{g}{\cos\theta_w}\, \eta_i \eta_j
\half \left(
\bold{N}_{i4} \bold{N^*}_{j4} -\bold{N}_{i3} \bold{N^*}_{j3}
-\sum_{k=1}^3 \bold{N}_{i,4+k} \bold{N^*}_{j,4+k} \right)\cr
\vb{22}
O^{\rm nnz}_{Rij}=- \frac{g}{\cos\theta_w} \half \left(
\bold{N^*}_{i4} \bold{N}_{j4} -\bold{N^*}_{i3} \bold{N}_{j3}
-\sum_{k=1}^3 \bold{N^*}_{i,4+k} \bold{N}_{j,4+k} \right)
\ea
\eeq

\subsubsection{Chargino--Neutralino--Charged Scalar}

\beq
\ba{l}\ds
O^{\rm cns}_{L ijk}= \eta_j \left[\vb{18}
{\bold R}^{S^{\pm}}_{k1}\left(
h_{E11} {\bold N^*}_{j5} {\bold V^*}_{i3}
+h_{E22} {\bold N^*}_{j6} {\bold V^*}_{i4}
+h_{E33} {\bold N^*}_{j7} {\bold V^*}_{i5}\right)\right. \cr
\vb{20}
\hskip 2.0cm
+{\bold R}^{S^{\pm}}_{k2}\left( 
- \frac{g}{\sqrt{2}} {\bold N^*}_{j2} {\bold V^*}_{i2}
-\frac{g'}{\sqrt{2}} {\bold N^*}_{j1} {\bold V^*}_{i2}
-g {\bold N^*}_{j4} {\bold V^*}_{i1}\right)
\cr
\vb{20}
\hskip 2.0cm
-{\bold R}^{S^{\pm}}_{k3} h_{E11} {\bold N^*}_{j3} {\bold V^*}_{i3}
-{\bold R}^{S^{\pm}}_{k4} h_{E22} {\bold N^*}_{j3} {\bold V^*}_{i4}
-{\bold R}^{S^{\pm}}_{k5} h_{E33} {\bold N^*}_{j3} {\bold V^*}_{i5}
\cr
\vb{20}\left.\vb{18}
\hskip 2.0cm
-{\bold R}^{S^{\pm}}_{k6} g' \sqrt{2} {\bold N^*}_{j1} {\bold V^*}_{i3}
-{\bold R}^{S^{\pm}}_{k7} g' \sqrt{2} {\bold N^*}_{j1} {\bold V^*}_{i4}
-{\bold R}^{S^{\pm}}_{k8} g' \sqrt{2} {\bold N^*}_{j1} {\bold V^*}_{i5}
\right]
\ea
\eeq

\beq
\ba{l}\ds
O^{\rm cns}_{R ijk}=\eta_i \left[\vb{18}
{\bold R}^{S^{\pm}}_{k1}\left( 
\frac{g}{\sqrt{2}} {\bold N}_{j2} {\bold U}_{i2}
+\frac{g'}{\sqrt{2}} {\bold N}_{j1} {\bold U}_{i2}
-g {\bold N}_{j3} {\bold U}_{i1}\right)\right.
\cr
\vb{20}\hskip 2.0cm
+{\bold R}^{S^{\pm}}_{k3}\left( 
\frac{g}{\sqrt{2}} {\bold N}_{j2} {\bold U}_{i3}
+\frac{g'}{\sqrt{2}} {\bold N}_{j1} {\bold U}_{i3}
-g {\bold N}_{j5} {\bold U}_{i1}\right)
\cr
\vb{20}\hskip 2.0cm
+{\bold R}^{S^{\pm}}_{k4}\left( 
\frac{g}{\sqrt{2}} {\bold N}_{j2} {\bold U}_{i4}
+\frac{g'}{\sqrt{2}} {\bold N}_{j1} {\bold U}_{i4}
-g {\bold N}_{j6} {\bold U}_{i1}\right)
\cr
\vb{20}\hskip 2.0cm
+{\bold R}^{S^{\pm}}_{k5}\left( 
\frac{g}{\sqrt{2}} {\bold N}_{j2} {\bold U}_{i5}
+\frac{g'}{\sqrt{2}} {\bold N}_{j1} {\bold U}_{i5}
-g {\bold N}_{j7} {\bold U}_{i1}\right)
\cr
\vb{20}\hskip 2.0cm
+{\bold R}^{S^{\pm}}_{k6} h_{E11}
\left( {\bold N}_{j5} {\bold U}_{i2} - {\bold N}_{j3} {\bold U}_{i3}\right)
+{\bold R}^{S^{\pm}}_{k7} h_{E22}
\left( {\bold N}_{j6} {\bold U}_{i2} - {\bold N}_{j3} {\bold U}_{i4}\right)
\cr
\vb{20}\hskip 2.0cm
\left. \vb{18}
+{\bold R}^{S^{\pm}}_{k8} h_{E33}
\left( {\bold N}_{j7} {\bold U}_{i2} - {\bold N}_{j3} {\bold U}_{i5}\right)
\right]
\ea
\eeq

\beq
O^{\rm ncs}_{L ijk}= \left( O^{\rm ncs}_{R jik} \right)^*
\hskip 5mm ; \hskip 5mm
O^{\rm cns}_{R ijk}=  \left( O^{\rm ncs}_{L jik} \right)^*
\eeq

\subsubsection{Neutralino--Neutralino--Scalar}

\beq
\ba{l}\ds
O^{\rm nnh}_{L ijk}=\eta_j\, \frac{1}{2} \left[\vb{18}
{\bold R}^{S^0}_{k1}\left(
-g\, {\bold N^*}_{i2} {\bold N^*}_{j3}
+g'\, {\bold N^*}_{i1} {\bold N^*}_{j3}
-g\, {\bold N^*}_{j2} {\bold N^*}_{i3}
+g'\, {\bold N^*}_{j1} {\bold N^*}_{i3} \right)\right.
\cr
\vb{22}\hskip 2.0cm\ds
+{\bold R}^{S^0}_{k2}\left(
+g\, {\bold N^*}_{i2} {\bold N^*}_{j4}
-g'\, {\bold N^*}_{i1} {\bold N^*}_{j4}
+g\, {\bold N^*}_{j2} {\bold N^*}_{i4}
-g'\, {\bold N^*}_{j1} {\bold N^*}_{i4} \right)
\cr
\vb{22}\hskip 2.0cm\ds
+{\bold R}^{S^0}_{k3}\left(
-g\, {\bold N^*}_{i2} {\bold N^*}_{j5}
+g'\, {\bold N^*}_{i1} {\bold N^*}_{j5}
-g\, {\bold N^*}_{j2} {\bold N^*}_{i5}
+g'\, {\bold N^*}_{j1} {\bold N^*}_{i5} \right)
\cr
\vb{22}\hskip 2.0cm\ds
+{\bold R}^{S^0}_{k4}\left(
-g\, {\bold N^*}_{i2} {\bold N^*}_{j6}
+g'\, {\bold N^*}_{i1} {\bold N^*}_{j6}
-g\, {\bold N^*}_{j2} {\bold N^*}_{i6}
+g'\, {\bold N^*}_{j1} {\bold N^*}_{i6} \right)
\cr
\vb{22}\hskip 2.0cm
\left. \vb{18}\ds
+{\bold R}^{S^0}_{k5}\left(
-g\, {\bold N^*}_{i2} {\bold N^*}_{j7}
+g'\, {\bold N^*}_{i1} {\bold N^*}_{j7}
-g\, {\bold N^*}_{j2} {\bold N^*}_{i7}
+g'\, {\bold N^*}_{j1} {\bold N^*}_{i7} \right)\right]
\cr
\vb{22}
O^{\rm nnh}_{R ijk}= \left( O^{\rm nnh}_{L jik} \right)^*
\ea
\eeq

\subsubsection{Neutralino--Neutralino--Pseudo Scalar}

\beq
\ba{l}\ds
O^{\rm nna}_{L ijk}= 
- \eta_j\, \frac{1}{2} \left[\vb{18}
{\bold R}^{P^0}_{k1}\left(
-g\, {\bold N^*}_{i2} {\bold N^*}_{j3}
+g'\, {\bold N^*}_{i1} {\bold N^*}_{j3}
-g\, {\bold N^*}_{j2} {\bold N^*}_{i3}
+g'\, {\bold N^*}_{j1} {\bold N^*}_{i3} \right)\right.
\cr
\vb{22}\hskip 2.0cm\ds
+{\bold R}^{P^0}_{k2}\left(
+g\, {\bold N^*}_{i2} {\bold N^*}_{j4}
-g'\, {\bold N^*}_{i1} {\bold N^*}_{j4}
+g\, {\bold N^*}_{j2} {\bold N^*}_{i4}
-g'\, {\bold N^*}_{j1} {\bold N^*}_{i4} \right)
\cr
\vb{22}\hskip 2.0cm\ds
+{\bold R}^{P^0}_{k3}\left(
-g\, {\bold N^*}_{i2} {\bold N^*}_{j5}
+g'\, {\bold N^*}_{i1} {\bold N^*}_{j5}
-g\, {\bold N^*}_{j2} {\bold N^*}_{i5}
+g'\, {\bold N^*}_{j1} {\bold N^*}_{i5} \right)
\cr
\vb{22}\hskip 2.0cm\ds
+{\bold R}^{P^0}_{k4}\left(
-g\, {\bold N^*}_{i2} {\bold N^*}_{j6}
+g'\, {\bold N^*}_{i1} {\bold N^*}_{j6}
-g\, {\bold N^*}_{j2} {\bold N^*}_{i6}
+g'\, {\bold N^*}_{j1} {\bold N^*}_{i6} \right)
\cr
\vb{22}\hskip 2.0cm
\left. \vb{18}\ds
+{\bold R}^{P^0}_{k5}\left(
-g\, {\bold N^*}_{i2} {\bold N^*}_{j7}
+g'\, {\bold N^*}_{i1} {\bold N^*}_{j7}
-g\, {\bold N^*}_{j2} {\bold N^*}_{i7}
+g'\, {\bold N^*}_{j1} {\bold N^*}_{i7} \right)\right]
\cr
\vb{22}
O^{\rm nna}_{R ijk}=-\left( O^{\rm nna}_{L jik} \right)^*
\ea
\eeq

\ni
The factors $\eta_i$ are the signs one has to include if we consider
${\bold N}$, ${\bold U}$ and ${\bold V}$ as real matrices and the mass of
the fermion $i$ is negative.

\subsubsection{Neutralino--Up Quark--Up Squark}

\beq
O^{\rm uns}_{L ijk} = \frac{4}{3}(\frac{g}{\sqrt{2}}) 
        \tan\theta_W {\bold N^*}_{j1} {\bold R}^{\tilde u^*}_{k,m+3} 
        {\bold R}^{u}_{{\bold R}i,m}
        - (h_u)_{ml}{\bold R}^{\tilde u^*}_{k,m}
          {\bold R}^{u}_{{\bold R}i,l}{\bold N^*}_{j4}
\eeq

\beq
O^{\rm uns}_{R ijk} = -(\frac{g}{\sqrt{2}})( {\bold N}_{j2}
    +    \frac{1}{3}\tan\theta_W {\bold N}_{j1}) 
        {\bold R}^{\tilde u^*}_{k,m} 
        {\bold R^*}^{u}_{{\bold L}m,i}
        - (h_u^*)_{ml}{\bold R}^{\tilde u^*}_{k+3,l}
          {\bold R^*}^{u}_{{\bold L}m,i}{\bold N}_{j4}
\eeq
and, 
\beq
O^{\rm nus}_{L ijk}= \left( O^{\rm uns}_{R jik} \right)^*
\hskip 5mm ; \hskip 5mm
O^{\rm nus}_{R ijk}= \left( O^{\rm uns}_{L jik} \right)^*
\eeq

\subsubsection{Neutralino--Down Quark--Down Squark}

\beq
O^{\rm dns}_{L ijk} = -\frac{2}{3}(\frac{g}{\sqrt{2}}) 
        \tan\theta_W {\bold N^*}_{j1} {\bold R}^{\tilde d^*}_{k,m+3} 
        {\bold R}^{d}_{{\bold R}i,m}
        - (h_d)_{ml}{\bold R}^{\tilde d^*}_{k,m}
          {\bold R}^{d}_{{\bold R}i,l}{\bold N^*}_{j3}
\eeq

\beq
O^{\rm dns}_{R ijk} = (\frac{g}{\sqrt{2}})( {\bold N}_{j2}
       - \frac{1}{3}\tan\theta_W {\bold N}_{j1} )
        {\bold R}^{\tilde d^*}_{k,m} 
        {\bold R^*}^{d}_{{\bold L}m,i}
        - (h_d^*)_{ml}{\bold R}^{\tilde d^*}_{k,l+3}
          {\bold R^*}^{d}_{{\bold L}m,i}{\bold N}_{j3}
\eeq
and, 
\beq
O^{\rm nds}_{L ijk}= \left( O^{\rm dns}_{R jik} \right)^*
\hskip 5mm ; \hskip 5mm
O^{\rm nds}_{R ijk}= \left( O^{\rm dns}_{L jik} \right)^*
\eeq

\subsection{The Neutral Scalar Couplings}

To evaluate the tadpoles we need the couplings of
the neutral scalars  with all the fields in the model. These couplings
are easier to write in the unrotated basis. The couplings for the mass
eigenstates can always be obtained by appropriate multiplication by
the rotation matrices. As an example, and to fix the notation
(repeated indices are understood to be summed unless otherwise stated), the
couplings of three neutral scalars in the two basis will be related by
\beq
g_{ijk}^{S^0S^0S^0}={\bold R}^{S^0}_{ip}\,
{\bold R}^{S^0}_{jq}\, {\bold R}^{S^0}_{kr}\,
g_{pqr}^{S'^0S'^0S'^0}
\eeq
Sometimes we will also use partially rotated couplings, for instance
\beq
g_{ijk}^{S'^0S^0S^0}=
{\bold R}^{S^0}_{jq}\, {\bold R}^{S^0}_{kr}\,
g_{iqr}^{S'^0S'^0S'^0}
\eeq
in an obvious notation. These couplings are defined as follows
\beqa
g_{ijk}^{S^0S^0S^0}&=& \frac{\partial^3 {\cal L}}
{ \partial S^0_i \partial S^0_j \partial S^0_k}\cr
\vb{24}
g_{ijk}^{S'^0S'^0S'^0}&=&\frac{\partial^3 {\cal L}}
{ \partial S'^0_i \partial S'^0_j \partial S'^0_k}
\eeqa

\subsubsection{Neutral Scalar--Neutral Scalar--Neutral Scalar}

\beq
g_{ijk}^{S'^0S'^0S'^0}=-\smallfrac{1}{4}\, \left(g^2+g'^2\right)\,
u_m \left( \hat \delta_{mi}\, \hat \delta_{jk} 
+\hat \delta_{mj}\, \hat \delta_{ik} +
\hat \delta_{mk}\, \hat \delta_{ij} \right)
\eeq

\noindent
where we have defined 
\beq
u_m \equiv (v_d,v_u,v_1,v_2,v_3) \hskip 10mm ; \hskip 10mm
\hat \delta_{ij}\equiv \hbox{diag}(+,-,+,+,+)
\eeq
For future reference we also define 
\beq
v_m\equiv (v_1,v_2,v_3)
\eeq
while $\delta_{ij}$ without the {\it hat} is the usual Kronecker delta.

\subsubsection{Scalar--Pseudo Scalar--Pseudo Scalar}

\beq
g_{ijk}^{S'^0P'^0P'^0}=-\smallfrac{1}{4}\, \left(g^2+g'^2\right)\,
 u_m \, \hat \delta_{mi} \hat \delta_{jk}
\eeq

\subsubsection{Scalar--Charged Scalar--Charged Scalar}

We define
\beq
g_{ijk}^{S'^0S'^+S'^-}=\left(
\matrix{
g_{iHH}^{S'^0S'^+S'^-}&g_{iHL}^{S'^0S'^+S'^-}&g_{iHR}^{S'^0S'^+S'^-}\cr
\vb{18}
\left(g_{iHL}^{S'^0S'^+S'^-}\right)^{\dagger}
&g_{iLL}^{S'^0S'^+S'^-}&g_{iLR}^{S'^0S'^+S'^-}\cr
\vb{18}
\left(g_{iHR}^{S'^0S'^+S'^-}\right)^{\dagger}
&\left(g_{iLR}^{S'^0S'^+S'^-}\right)^{\dagger}&g_{iRR}^{S'^0S'^+S'^-}
}
\right)
\eeq
where
\beqa
\left(g_{iHH}^{S'^0S'^+S'^-}\right)_{jk}&=&
\smallfrac{1}{4}\, g^2\left[ \vb{16}
-v_u \left( 
\delta_{i1}\, \delta_{j1}\, \delta_{k2} +
\delta_{i2}\, \delta_{j1}\, \delta_{k1} +
\delta_{i1}\, \delta_{j2}\, \delta_{k1} +
\delta_{i2}\, \delta_{j2}\, \delta_{k2}\right) \right. \cr
\vb{16}
&& \hskip 12mm  -v_d \left( 
\delta_{i1}\, \delta_{j1}\, \delta_{k1} +
\delta_{i1}\, \delta_{j2}\, \delta_{k2} +
\delta_{i2}\, \delta_{j1}\, \delta_{k2} +
\delta_{i2}\, \delta_{j2}\, \delta_{k1}\right)  \cr
\vb{18}
&&\hskip 12mm\left. \vb{16}
 + v_m\, \delta_{i-2,m}\, \left( \delta_{j1}\, \delta_{k1} 
- \delta_{j2}\, \delta_{k2} \right) \right] \cr
\vb{18}
&&\hskip 12mm
-\smallfrac{1}{4}\, g'^2\, u_m\, \hat \delta_{im}\, \hat \delta_{jk} \cr
\vb{18}
&&\hskip 12mm
-\smallfrac{1}{2}\,  v_m\, \delta_{j1}\, \delta_{k1}\, 
\left( h_E h_E^{\dagger} +h_E^* h_E^T
\right)_{m,i-2}
\eeqa
\beqa
\left(g_{iHL}^{S'^0S'^+S'^-}\right)_{jk}&=&
-\smallfrac{1}{4}\, g^2\,  \left[\vb{16}
\delta_{i-2,k}\,\left( v_d\,
 \delta_{j1} +v_u\, \delta_{j2} \right)
+v_m\, \delta_{ij}\, \delta_{mk} \right] + \smallfrac{1}{2}\, 
v_m \left( h_E^* h_E^T \right)_{mk}\, \delta_{i1}\, \delta_{j1} \cr
\vb{18}
&&
+\smallfrac{1}{2}\, v_d\, \delta_{j1}\, \left(h_E^* h_E^T \right)_{i-2,k}
\eeqa
\beqa
\left(g_{iHR}^{S'^0S'^+S'^-}\right)_{jk}&=&
\smallfrac{1}{\sqrt{2}}\, \epsilon_m\, \left(h_E^*\right)_{mk}\,
\left(\delta_{i1}\, \delta_{j2} + \delta_{i2}\, \delta_{j1} \right)
+\smallfrac{1}{\sqrt{2}}\,  \left(A_E^*\right)_{i-2,k}\,
\delta_{j1}\cr
\vb{18}
&&
+\smallfrac{1}{\sqrt{2}}\, \mu\,  \left(h_E^*\right)_{i-2,k}\, \delta_{j2}
\eeqa
\beqa
\left(g_{iLL}^{S'^0S'^+S'^-}\right)_{jk}&=&
\smallfrac{1}{4}\, \left(g^2-g'^2\right)\, u_m\, \hat \delta_{im}\, 
\delta_{jk} - \smallfrac{1}{4}\, g^2 \, v_m \left( \delta_{i-2,j}\,
\delta_{mk} + \delta_{i-2,k}\, \delta_{mj} \right) \cr
\vb{18}
&&
-\left(h_E^*h_E^T\right)_{jk}\, v_d\, \delta_{i1}
\eeqa
\beq
\left(g_{iLR}^{S'^0S'^+S'^-}\right)_{jk}=
-\smallfrac{1}{\sqrt{2}}\, \delta_{i1}\, \left(A_E^*\right)_{jk} 
+\smallfrac{1}{\sqrt{2}}\, \mu\, \delta_{i2}\, \left(h_E^*\right)_{jk} 
\eeq
\beqa
\left(g_{iRR}^{S'^0S'^+S'^-}\right)_{jk}&=&
\smallfrac{1}{2}\, g'^2\, u_m\, \hat \delta_{im}\, \delta_{jk} 
-v_d\, \delta_{i1}\, \left(h_E^Th_E^*)\right)_{jk} \cr
\vb{18}
&&
-\smallfrac{1}{2}\, v_m\, \left[
\left(h_E^*\right)_{i-2,k}\, \left(h_E\right)_{mj}+
\left(h_E\right)_{i-2,j}\, \left(h_E^*\right)_{mk}\right]
\eeqa

\subsubsection{Scalar--Up Squarks--Up Squarks}

With the definition
\beq
{\cal L}= g_{ijk}^{S'^0\widetilde{u}'\widetilde{u}'^*}\,
S'^0_i\, \widetilde{u}'_j\, \widetilde{u}'^*_k + \cdots
\eeq
we get
\beq
g_{ijk}^{S'^0\widetilde{u}'\widetilde{u}'^*}=\left(
\matrix{
g_{iLL}^{S'^0\widetilde{u}'\widetilde{u}'^*}&
g_{iLR}^{S'^0\widetilde{u}'\widetilde{u}'^*}\cr
\vb{18}
g_{iRL}^{S'^0\widetilde{u}'\widetilde{u}'^*}&
g_{iRR}^{S'^0\widetilde{u}'\widetilde{u}'^*}\cr
}
\right)
\eeq
where
\beqa
g_{iLL}^{S'^0\widetilde{u}'\widetilde{u}'^*}
&=& u_m\, \hat \delta_{im} 
\left(-\smallfrac{1}{4}\, g^2 +\smallfrac{1}{12}\, g'^2
\right)
{\cal I}
-v_u \left( h_U h_U^{\dagger}\right)\, \delta_{i2}\cr
\vb{18}
g_{iLR}^{S'^0\widetilde{u}'\widetilde{u}'^*}
&=& - \smallfrac{1}{\sqrt{2}}\, \delta_{i2}\, A_U + 
\smallfrac{1}{\sqrt{2}}\, \mu\, h_U\, \delta_{i1}
-\smallfrac{1}{\sqrt{2}}\,  h_U\, \epsilon_m\, \delta_{i-2,m}
 \cr
\vb{18}
g_{iRL}^{S'^0\widetilde{u}'\widetilde{u}'^*}
&=& 
g_{iLR}^{H'^0\widetilde{u}'\widetilde{u}'^*}\cr
\vb{18}
g_{iRR}^{S'^0\widetilde{u}'\widetilde{u}'^*}
&=& -\smallfrac{1}{3} u_m\, \hat \delta_{im} \, g'^2 {\cal I} 
-v_u \left( h_U^T h_U^*\right)\, \delta_{i2}
\eeqa
where ${\cal I}$ is the unit $3\times 3$ matrix.

\subsubsection{Scalar--Down Squarks--Down Squarks}

With the definition
\beq
{\cal L}= g_{ijk}^{S'^0\widetilde{d}'\widetilde{d}'^*}\,
S'^0_i\, \widetilde{d}'_j\, \widetilde{d}'^*_k + \cdots
\eeq
we get
\beq
g_{ijk}^{S'^0\widetilde{d}'\widetilde{d}'^*}=\left(
\matrix{
g_{iLL}^{S'^0\widetilde{d}'\widetilde{d}'^*}&
g_{iLR}^{S'^0\widetilde{d}'\widetilde{d}'^*}\cr
\vb{18}
g_{iRL}^{S'^0\widetilde{d}'\widetilde{d}'^*}&
g_{iRR}^{S'^0\widetilde{d}'\widetilde{d}'^*}\cr
}
\right)
\eeq
where
\beqa
g_{iLL}^{S'^0\widetilde{d}'\widetilde{d}'^*}
&=& u_m \, \hat \delta_{im}\,
\left( \smallfrac{1}{4}\, g^2 +\smallfrac{1}{12}\, g'^2
\right)
{\cal I}
-v_d \left( h_D h_D^{\dagger}\right)\, \delta_{i1}\cr
\vb{18}
g_{iLR}^{S'^0\widetilde{d}'\widetilde{d}'^*}
&=& - \smallfrac{1}{\sqrt{2}}\, \delta_{i1}\, A_D + 
\smallfrac{1}{\sqrt{2}}\, \mu\, h_D\, \delta_{i2} \cr
\vb{18}
g_{iRL}^{S'^0\widetilde{d}'\widetilde{d}'^*}
&=& 
g_{iLR}^{S'^0\widetilde{d}'\widetilde{d}'^*}\cr
\vb{18}
g_{iRR}^{S'^0\widetilde{d}'\widetilde{d}'^*}
&=& \smallfrac{1}{6}\, u_m \, \hat \delta_{im} \, g'^2 {\cal I} 
-v_d \left( h_D^T h_D^*\right)\, \delta_{i1}
\eeqa

\subsubsection{Scalar--$W^+$--$W^-$}

With the definition
\beq
{\cal L}= g_{i}^{S'^0W^+W^-}\,
S'^0_i\, W^+\, W^- + \cdots
\eeq
we get
\beq
g_{i}^{S'^0W^+W^-}= g\, 
\frac{m_W}{v}\,
\left(v_d\, \delta_{i1}+v_u\, \delta_{i2}+ v_m\, \delta_{i-2,m}\right)
\label{wcoupl}
\eeq
where
\beq
v=\sqrt{v_d^2+v_u^2+v_1^2+v_2^2+v_3^2}
\eeq

\subsubsection{Scalar--$Z^0$--$Z^0$}

With the definition
\beq
{\cal L}= \half\, g_{i}^{S'^0Z^0Z^0}\,
S'^0_i\, Z^0 Z^0 + \cdots
\eeq
we get
\beq
g_{i}^{S'^0Z^0Z^0}= \frac{g}{\cos\theta_W}\, 
\frac{m_Z}{v}\,
\left(v_d\, \delta_{i1}+v_u\, \delta_{i2}+ v_m\, \delta_{i-2,m}\right)
\label{zcoupl}
\eeq

\subsubsection{Scalar--Quark--Quark}

With the definition
\beq
{\cal L}=  g_{ijk}^{S'^0\overline{u}u}\,
S'^0_i\, \overline{u}_j\, u_k +
g_{ijk}^{S'^0\overline{d}d}\,
S'^0_i\, \overline{d}_j\, d_k + \cdots
\eeq
we get
\beq
g_{ijk}^{S'^0\overline{u}u}=-\smallfrac{1}{\sqrt{2}}\, 
\left(h_U\right)_{jk} \, \delta_{i2}
\eeq
and
\beq
g_{ijk}^{S'^0\overline{d}d}=-\smallfrac{1}{\sqrt{2}}\, 
\left(h_D\right)_{jk} \, \delta_{i1}
\eeq

\subsubsection{Scalar--Chargino--Chargino {\rm and}
 Scalar--Neutralino--Neutralino}

With the definition
\beq
{\cal L}=  \overline{\chi_i^-}\left( O_{Lijk}^{\rm cch'}
+O_{Rijk}^{\rm cch'}
\right) \chi_j^-\,
S'^0_i +\frac{1}{2}\, 
\overline{\chi_i^0}\left( O_{Lijk}^{\rm nnh'} +O_{Rijk}^{\rm nnh'}
\right) \chi_j^0\,
S'^0_i 
\eeq
we have
\beqa
O^{\rm cch'}_{L ijk}&=& -\frac{\epsilon_j}{\sqrt{2}}\, 
\left[\vb{18} g 
\left(\vb{12}
{\bold V^*}_{i1} {\bold U^*}_{j2}\, \delta_{k1}
+{\bold V^*}_{i2} {\bold U^*}_{j1}\, \delta_{k2}
+{\bold V^*}_{i1} {\bold U^*}_{j3}\, \delta_{k3} \right. \right.\cr
\vb{16}
&&\left.\vb{12} \hskip 15mm
+{\bold V^*}_{i1} {\bold U^*}_{j4}\, \delta_{k4}
+{\bold V^*}_{i1} {\bold U^*}_{j5}\, \delta_{k5}
\right)\cr
\vb{18}
&& \hskip 12mm
+ \left( 
h_{E11} {\bold U^*}_{j3} {\bold V^*}_{i3}+
h_{E22} {\bold U^*}_{j4} {\bold V^*}_{i4}+
h_{E33} {\bold U^*}_{j5} {\bold V^*}_{i5}\right) \delta_{k1} \cr
\vb{18}
&&\left.\vb{18} \hskip 12mm
-\left(
h_{E11} {\bold U^*}_{j2} {\bold V^*}_{i3}\, \delta_{k3}+
h_{E22} {\bold U^*}_{j2} {\bold V^*}_{i4}\, \delta_{k4}+
h_{E33} {\bold U^*}_{j2} {\bold V^*}_{i5}\, \delta_{k5}\right) \right]\cr
\vb{18}
O^{\rm cch'}_{R ijk}&=&\left(O^{\rm cch'}_{L jik}\right)^*
\eeqa
and
\beqa
O^{\rm nnh'}_{L ijk}&=&\eta_j\, \frac{1}{2} 
\left(
-g\, {\bold N^*}_{i2} {\bold N^*}_{j3}
+g'\, {\bold N^*}_{i1} {\bold N^*}_{j3}
-g\, {\bold N^*}_{j2} {\bold N^*}_{i3}
+g'\, {\bold N^*}_{j1} {\bold N^*}_{i3} \right)\cr
\vb{18}
&&\hskip 10mm
\left(\delta_{k1}-\delta_{k2}+\delta_{k3}+\delta_{k4}
+\delta_{k5}\right)\cr
\vb{18}
O^{\rm nnh'}_{R ijk}&=&\left(O^{\rm nnh'}_{L jik}\right)^*
\eeqa

\section{Tadpoles}
\label{Ap:tadpoles}

\subsection{Gauge Boson and Ghost Tadpoles}

We will consider the gauge boson and ghost tadpoles in an arbitrary
$R_{\xi}$ gauge to show that the dependence on $\xi$ cancels out. We
will do it for any model.

\subsubsection{General $Z^0$ Boson Tadpole} 

We write down the tadpole contribution from the $Z^0$ for a general
theory with the coupling $g_{HZZ}$ to the higgs boson.

\vspace{1cm}

\begin{equation}
i T_Z= \half i\, g_{HZZ}\, \int \frac{d^d p}{(2\pi)^d}\,
G_{\mu}{}^{\mu} (p)
\end{equation}

\begin{picture}(0,0)
\put(0,-0.5){\includegraphics{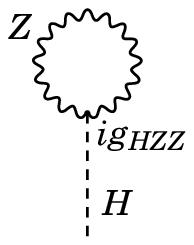}}
\end{picture}

\ni
where
\beq
G_{\mu}{}^{\mu}=(-i)\, \left[ \frac{4}{p^2-M_Z^2} -(1-\xi) 
\frac{p^2}{(p^2-M_Z^2) (p^2-\xi M_Z^2)} \right]
\eeq
and the factor $\half$ is a symmetry factor. Now we do some
transformations in the second term of the $Z^0$ propagator $G(\xi)$,
\beqa
G(\xi)&=&
(1-\xi) \frac{p^2}{p^2-M_Z^2}\, \frac{1}{p^2-\xi M_Z^2}\cr
\vb{22}
&=&\frac{1}{p^2-M_Z^2} 
- \xi\,  \frac{1}{p^2-\xi M_Z^2}
\eeqa
and therefore we can write
\beq
G_{\mu}{}^{\mu}=(-i)\, \left[ \frac{3}{p^2-M_Z^2} + \xi\,
\frac{1}{p^2-\xi M_Z^2} \right]
\eeq
Then
\beqa
iT_Z&=& \half i\, g_{HZZ}\, (-i)\, \frac{i}{16\pi^2}\,
\left[ 3 A_0(M_Z^2) + \xi\, A_0(\xi M_Z^2) \right]\cr
\vb{22}
&=& \frac{i}{16\pi^2}\, \half\, g_{HZZ}\, 
\left[ 3 A_0(M_Z^2) + \xi\, A_0(\xi M_Z^2) \right] 
\eeqa
where we have used the definition
\beq
\frac{i}{16\pi^2}\, A_0(m^2)\equiv
\int \frac{d^d p}{(2\pi)^d}\, \frac{1}{p^2-m^2}
\eeq
As $A_0(\xi m^2)$ grows for large $\xi$ as $\xi m^2$ we conclude that
$T_Z$ grows like $\xi^2$. This dependence has to cancel against other
diagrams. It is easy to realize that the Goldstone of the $Z^0$ will
not do it because, although its mass depend on $\xi$, its contribution
to the tadpole will only grow like $\xi$ because its coupling to $H$
does not depend on $\xi$. But the ghost coupling to $H$ does depend 
$\xi$ as we will see. 

\subsubsection{General $Z^0$ Ghost Tadpole}

Let us then calculate the tadpole of the ghost of the $Z^0$. We have

\vspace{1cm}

\begin{equation}
i T_{c_z}= (-1)\, i\, g_{H{c_z}{\overline{c}_z}}\, 
\int \frac{d^d p}{(2\pi)^d}\,
\frac{i}{p^2-\xi M_Z^2}
\end{equation}

\begin{picture}(0,0)
\put(0,-0.5){\includegraphics{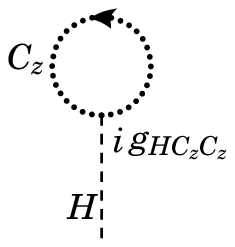}}
\end{picture}

\ni
where the factor $(-1)$ is because of the anti-commutative properties
of the ghosts. Using the definition of $A_0$ we get
\beq
i T_{c_z}= \frac{i}{16\pi^2}\,  g_{H{c_z}{\overline{c}_z}}\,
A_0(\xi M_Z^2)
\eeq
Adding the two contributions together we obtain
\beq
iT_Z + iT_{c_z}= \frac{i}{16\pi^2}\,
\left[\frac{3}{2}\, g_{HZZ}\, A_0(M_Z^2) + \left( \half g_{HZZ}\, \xi
+ g_{H{c_z}{\overline{c}_z}}\right)A_0(\xi M_Z^2) \right]
\eeq
We see that for the $\xi$ dependence to cancel one must have
\beq
\half g_{HZZ}\, \xi + g_{H{c_z}{\overline{c}_z}} =0
\label{zcancel}
\eeq
As we will show below this is true for the SM, MSSM and also for the Bilinear
R-Parity Model. Then the contribution from the $Z^0$ and neutral ghost
tadpoles is, for any model, gauge independent and given by
\beq
iT_Z + iT_{c_z}= \frac{i}{16\pi^2}\,
\frac{3}{2}\, g_{HZZ}\, A_0(M_Z^2) 
\label{tadz} 
\eeq

\subsubsection{The $W^{\pm}$ Boson and $c^{\pm}$ Ghost Tadpoles}

The calculation for the $W^{\pm}$ boson and charged ghosts is very similar.
The main differences are that the $W$ tadpole
does not have a factor $\half$ and that there are {\it two} ghosts for
the $W^{\pm}$. Therefore we have
\beqa
iT_W + iT_{c^+_W}+iT_{c^-_W}
&=& \frac{i}{16\pi^2}\, 
\left[\vb{18} 3\, g_{HWW}\, A_0(M_W^2) \right. \cr
&&
\left. \vb{18}
\hskip 10mm
+ \left(g_{HWW}\, \xi +  g_{H{c^+_W}{\overline{c^+}_W}}
+g_{H{c^-_W}{\overline{c^-}_W}}
\right)\, A_0(\xi M_W^2)\right]
\eeqa
We see that the $\xi$ dependence will cancel out if
\beq
g_{HWW}\, \xi +  g_{H{c^+_W}{\overline{c^+}_W}}
+g_{H{c^-_W}{\overline{c^-}_W}}
=0
\label{wcancel}
\eeq
We will show below that this is true in general. 
Then the contribution from the $W^{\pm}$ and charged ghost
tadpoles is, for any model, gauge independent and given by
\beq
iT_W + iT_{c^+_W}+iT_{c^-_W}
= \frac{i}{16\pi^2}\,
3\, g_{HWW}\, A_0(M_W^2) 
\label{tadw} 
\eeq

\subsubsection{The Standard Model}

Now lets us see us the cancellation occurs in the Standard Model (SM).
The relevant couplings for the $Z^0$ are
\beqa
g_{HZZ}&=& \frac{g}{\cos \theta_W}\, M_Z \cr
\vb{22}
g_{H{c_z}{\overline{c}_z}}&=& - \frac{g}{2 \cos \theta_W}\, \xi\, M_Z
\eeqa
and we immediatly see that Eq.(\ref{zcancel}) is verified. For the
$W^{\pm}$ we have
\beqa
g_{HWW}&=& g\, M_W \cr
\vb{22}
g_{H{c^+_W}{\overline{c^+}_W}}&=& - \frac{g}{2}\, \xi\, M_W\cr
\vb{22}
g_{H{c^-_W}{\overline{c^-}_W}}&=& - \frac{g}{2}\, \xi\, M_W
\eeqa
satisfying Eq.(\ref{wcancel}).

\subsubsection{Bilinear R-Parity Model}

In the bilinear R-parity model the relevant couplings are 
\beqa
g_{i}^{S'^0{c^+_W}{\overline{c^+}_W}}&=&-\frac{g}{2}\, \xi\,
\frac{m_W}{v}\,
\left(v_d\, \delta_{i1}+v_u\, \delta_{i2}+ v_m\,
\delta_{i-2,m}\right)\cr
\vb{18}
g_{i}^{S'^0{c^-_W}{\overline{c^-}_W}}&=&-\frac{g}{2}\, \xi\,
\frac{m_W}{v}\,
\left(v_d\, \delta_{i1}+v_u\, \delta_{i2}+ v_m\,
\delta_{i-2,m}\right)
\label{wghost}
\eeqa
and
\beq
g_{i}^{S'^0{c_z}{\overline{c}_z}}=-\frac{g}{2\cos\theta_W}\, \xi\,
\frac{m_Z}{v}\,
\left(v_d\, \delta_{i1}+v_u\, \delta_{i2}+ v_m\,
\delta_{i-2,m}\right)
\label{zghost}
\eeq
Then using Eqs.(\ref{wcoupl},\ref{zcoupl},\ref{wghost},\ref{zghost}) in 
Eqs.(\ref{wcancel},\ref{zcancel}) we see that the same
cancellation occurs.

\subsection{General Tadpole Expressions}

After showing the gauge invariance of the gauge boson tadpoles
together with their ghosts we give now the general tadpole in a compact
form. We will write them for the unrotated neutral Higgs $H'^0$ because
that is what is needed for substitution into Eq.(\ref{eq:tadpoles}).
The general form can be written as ($X=W^{\pm},Z^0,
S^{\pm},H^0,A^0,\widetilde{u},\widetilde{d},u,d$), 
\beq
T_{H_i'^0}^X= \frac{1}{16\pi^2}\, P^X_i
\eeq
where
\beqa
P_i^W \hskip -2mm&=&\hskip -2mm
\phantom{-} 3\, g_i^{S'^0W^+W^-}\, A_0(M_W^2)\cr
\vb{20}
P_i^Z \hskip -2mm&=&\hskip -2mm
\phantom{-} \frac{3}{2}\, g_i^{S'^0Z^0Z^0}\, A_0(M_Z^2)\cr
\vb{20}
P_i^{S^{\pm}} \hskip -2mm&=&\hskip -2mm
- \sum_{k=1}^8{}^{\prime}\, g_{ikk}^{S'^0S^+S^-}\, A_0(m_k^2)\cr
\vb{20}
P_i^{S^0} \hskip -2mm&=&\hskip -2mm
- \sum_{k=1}^5\, \frac{1}{2}\, g_{ikk}^{S'^0S^0S^0}\, A_0(m_k^2)\cr
\vb{20}
P_i^{P^0} \hskip -2mm&=&\hskip -2mm
- \sum_{k=1}^5{}^{\prime}\, \frac{1}{2}\, g_{ikk}^{S'^0P^0P^0}\, A_0(m_k^2)\cr
\vb{20}
P_i^{\, \widetilde{u}} \hskip -2mm&=&\hskip -2mm
- \sum_{k=1}^6\, 3\, g_{ikk}^{S'^0\widetilde{u}\widetilde{u}^*}\, A_0(m_k^2)\cr
\vb{20}
P_i^{\, \widetilde{d}} \hskip -2mm&=&\hskip -2mm
- \sum_{k=1}^6\, 3\, g_{ikk}^{S'^0\widetilde{d}\widetilde{d}^*}\, A_0(m_k^2)\cr
\vb{20}
P_i^{\chi^{\pm}} \hskip -2mm&=&\hskip -2mm
- \sum_{k=1}^5\, (-1)\, 
\frac{O_{Lkki}^{\rm cch'}+O_{Rkki}^{\rm cch'}}{2}\, 4 m_k\, 
A_0(m_k^2)\cr
\vb{20}
P_i^{\chi^0} \hskip -2mm&=&\hskip -2mm
- \sum_{k=1}^7\, (-\frac{1}{2})\, 
\frac{O_{Lkki}^{\rm nnh'}+O_{Rkki}^{\rm nnh'}}{2}\, 4 m_k\,
A_0(m_k^2)\cr
\vb{20}
P_i^{u} \hskip -2mm&=&\hskip -2mm
- \sum_{k=1}^3\, (-3)\, g_{ikk}^{H'^0\overline{u}u}\, 4 m_k\, A_0(m_k^2)\cr
\vb{20}
P_i^{d} \hskip -2mm&=&\hskip -2mm
- \sum_{k=1}^3\, (-3)\, g_{ikk}^{H'^0\overline{d}d}\, 4 m_k\, A_0(m_k^2)
\eeqa
where $\sum{}^{\prime}$ means that we sum over all fields {\it except}
for the goldstone boson. As explained in section \ref{gauinv} the
contribution of the goldstones is added to the self--energies to
achieve gauge invariance.

\section{One Loop Self--Energies}
\label{OneLoopSelfEnergies}

In this section we write down the contribution of the several self
energy diagrams in the $\xi=1$ gauge.

\subsection{The $W$ and $Z$ Loops}

The contribution of the $W$ and $Z$ loops to the functions $\Sigma^V$
and $\Pi^V$ can be written in the form ($X=W,Z$),

\beqa
\Sigma^V_{ij}&=& -\frac{1}{16\pi^2}\, \sum_k
F_{ijk}^X\,  B_1(p^2,m^2_k,m^2_X)\cr
\vb{20}
\Pi^V_{ij}&=& -\frac{1}{16\pi^2}\, \sum_k
G_{ijk}^X\, m_k\, B_0(p^2,m^2_k,m^2_X)
\eeqa
with
\beqa
F^W_{ijk}&=& 
\phantom{-}2 \left(O^{\rm ncw}_{L jk} O^{\rm cnw}_{L ki} +
O^{\rm ncw}_{R jk} O^{\rm cnw}_{R ki}\right)\cr
\vb{20}
G^W_{ijk}&=& 
-4 \left(O^{\rm ncw}_{L jk} O^{\rm cnw}_{R ki} +
O^{\rm ncw}_{R jk} O^{\rm cnw}_{L ki}\right)
\eeqa
and 
\beqa
F^Z_{ijk}&=&
\phantom{-2}\left(O^{\rm nnz}_{L jk} O^{\rm nnz}_{L ki} +
O^{\rm nnz}_{R jk} O^{\rm nnz}_{R ki}\right)\cr
\vb{20}
G^Z_{ijk}&=&
-2 \left(O^{\rm nnz}_{L jk} O^{\rm nnz}_{R ki} +
O^{\rm nnz}_{R jk} O^{\rm nnz}_{L ki}\right)
\eeqa

\subsection{The Scalar Loops}

All the scalar contributions can be written in the form
($X=S^{\pm},S^0,P^0,\widetilde{u},\widetilde{d}$),

\beqa
\Sigma^V_{ij}&=& -\frac{1}{16\pi^2}\, \sum_r\, \sum_k
F^X_{ijkr}\, B_1(p^2,m^2_k,m^2_r)\cr
\vb{20}
\Pi^V_{ij}&=& -\frac{1}{16\pi^2}\, \sum_r\, \sum_k
G^X_{ijkr}\, m_k\, B_0(p^2,m^2_k,m^2_r)
\eeqa
with
\beqa
F^{S^{\pm}}_{ijkr}\hskip -1mm&=& \hskip -1mm
\phantom{-\half} \left(O^{\rm ncs}_{R jkr} O^{\rm cns}_{L kir} +
O^{\rm ncs}_{L jkr} O^{\rm cns}_{R kir}\right)\cr
\vb{20}
G^{S^{\pm}}_{ijkr}\hskip -1mm&=& \hskip -1mm
\phantom{-\half}\left(O^{\rm ncs}_{L jkr} O^{\rm cns}_{L kir} +
O^{\rm ncs}_{R jkr} O^{\rm cns}_{R kir}\right)
\eeqa
\beqa
F^{S^0}_{ijkr}\hskip -1mm&=& \hskip -1mm
\phantom{-}\half
 \left(O^{\rm nnh}_{L jkr} O^{\rm nnh}_{L kir} +
O^{\rm nnh}_{R jkr} O^{\rm nnh}_{R kir}\right)\cr
\vb{20}
G^{S^0}_{ijkr}\hskip -1mm&=& \hskip -1mm
\phantom{-}\half
\left(O^{\rm nnh}_{L jkr} O^{\rm nnh}_{R kir} +
O^{\rm nnh}_{R jkr} O^{\rm nnh}_{L kir}\right)
\eeqa
\beqa
F^{P^0}_{ijkr}\hskip -1mm&=& \hskip -1mm
-\half
 \left(O^{\rm nna}_{R jkr} O^{\rm nna}_{L kir} +
O^{\rm nna}_{L jkr} O^{\rm nna}_{R kir}\right)\cr
\vb{20}
G^{P^0}_{ijkr}\hskip -1mm&=& \hskip -1mm
-\half
\left(O^{\rm nna}_{L jkr} O^{\rm nna}_{L kir} +
O^{\rm nna}_{R jkr} O^{\rm nna}_{R kir}\right)
\eeqa
\beqa
F^{\widetilde{u}}_{ijkr}\hskip -1mm&=& \hskip -1mm
\phantom{-\half}
 \left(O^{\rm nus}_{R jkr} O^{\rm uns}_{L kir} +
O^{\rm nus}_{L jkr} O^{\rm uns}_{R kir}\right)\cr
\vb{20}
G^{\widetilde{u}}_{ijkr}\hskip -1mm&=& \hskip -1mm
\phantom{-\half}
\left(O^{\rm nus}_{L jkr} O^{\rm uns}_{L kir} +
O^{\rm nus}_{R jkr} O^{\rm uns}_{R kir}\right)
\eeqa
\beqa
F^{\widetilde{d}}_{ijkr}\hskip -1mm&=& \hskip -1mm
\phantom{-\half}
 \left(O^{\rm nds}_{R jkr} O^{\rm dns}_{L kir} +
O^{\rm nds}_{L jkr} O^{\rm dns}_{R kir}\right)\cr
\vb{20}
G^{\widetilde{d}}_{ijkr}\hskip -1mm&=& \hskip -1mm
\phantom{-\half}
\left(O^{\rm nds}_{L jkr} O^{\rm dns}_{L kir} +
O^{\rm nds}_{R jkr} O^{\rm dns}_{R kir}\right)
\eeqa

\end{document}